\documentclass[conference]{IEEEtran}
\usepackage{cite}
\usepackage{amsmath,amssymb,amsfonts}
\usepackage{array,multirow,makecell}
\usepackage{graphicx}
\usepackage{textcomp}
\usepackage{xcolor}
\usepackage{xspace}
\usepackage[hyphens]{url}
\usepackage{changepage}
\usepackage{tikz}
\usepackage[utf8]{inputenc}
\usepackage[section]{algorithm}
\usepackage{algorithmicx}
\usepackage{algpseudocode}
\usepackage{wrapfig}
\usepackage[export]{adjustbox}

\def\BibTeX{{\rm B\kern-.05em{\sc i\kern-.025em b}\kern-.08em
    T\kern-.1667em\lower.7ex\hbox{E}\kern-.125emX}}

\graphicspath{{./Figs/}}

\definecolor{mygreen}{RGB}{0,90,0}

\newcommand{\fig}[1]{Figure~\ref{#1}\xspace}
\newcommand{\tab}[1]{Table~\ref{#1}\xspace}

\usepackage{xcolor,colortbl}
\definecolor{green}{rgb}{0.1,0.1,0.1}
\definecolor{babyblue}{rgb}{0.54, 0.81, 0.94}
\definecolor{beaublue}{rgb}{0.74, 0.83, 0.9}
\definecolor{bleudefrance}{rgb}{0.19, 0.55, 0.91}
\definecolor{darkseagreen}{rgb}{0.56, 0.74, 0.56}
\definecolor{azure}{rgb}{0.94, 1.0, 1.0}
\definecolor{beige}{rgb}{0.96, 0.96, 0.86}
\definecolor{bubbles}{rgb}{0.91, 1.0, 1.0}
\definecolor{eggshell}{rgb}{0.94, 0.92, 0.84}
\definecolor{gainsboro}{rgb}{0.86, 0.86, 0.86}
\definecolor{honeydew}{rgb}{0.94, 1.0, 0.94}

\newcommand{\intel}{\cellcolor{azure}}  %{0.9}
\newcommand{\amd}  {\cellcolor{honeydew}}

\newcommand{\ballnumber}[1]{\tikz[baseline=(myanchor.base)] \node[circle,fill=blue,inner sep=1pt] (myanchor) {\color{white}\bfseries\footnotesize #1};}

\title{Harvesting L2 Caches in Server Processors} 
\author{Majid Jalili, and Mattan Erez \\ The University of Texas at Austin\\\{majid,mattan.erez\}@utexas.edu}
\begin{document}
\maketitle
\thispagestyle{plain}
\pagestyle{plain}

%%%%%% -- PAPER CONTENT STARTS-- %%%%%%%%

\begin{abstract}
We make three observations in modern processors: (1) LLC capacity is getting larger (up to 1GB); (2) core counts are increasing (up to 128 cores), accumulating a more significant amount of private L2 cache capacity on the chip; and (3) overall processor utilization in the cloud remains very low despite many efforts, leaving many large private caches unused. To enable better use of these beefy processors, we propose to open up a logical path for LLC evictions to unused private caches. In other words, instead of writing LLC evictions back to slow and busy main memory, we send some of them that are still alive up to idle L2 caches to avoid unnecessary long and costly main memory. Our scheme takes the importance of applications (user-facing vs. background), and system load into account to provide each application with a fair share of idle resources. Our results show that we can improve system performance by up to 2$\times$ (geomean of 10\%) for single-application runs. Also, for mixes of user-facing and background jobs, our scheme improves the P99 latency of user-facing tasks by up to 32\% (geomean of 15\%), and the IPC of background jobs by up to 50\% (geomean of 10\%).
\end{abstract}

\section{Introduction}
\label{sec:intro}
CPU manufacturers are increasing the L2/LLC sizes and number of cores to respond to the ever-increasing demand for computation. For instance, comparing two high-end Intel processors (Xeon Platinum 8180 and Xeon Platinum 8380), we observe that total cache capacity has increased from 66.5MB to 110MB. AMD processors also follow a similar trends: EPYC-7773X has 800MB on-chip memory with 64 cores. At the same time, these CPUs are operating at very low utilization, 40\% at Azure \cite{resourceCentral}, and 20-50\% at Alibaba \cite{AlibabaUtil}. This shift toward deeper and larger caches combined with low utilization opens opportunities for novel cache management mechanisms. 

We propose L2 Harvester (L2H), a completely software-transparent scheme built on top of the current rigid memory hierarchy that harvests idle cache resources, reducing the average load latency by up to 30\%. L2H moves a fraction of LLC capacity/conflict evictions to unused private L2 caches instead of writing them back to main memory. Thus, later L2 misses can find data in other L2 caches, reducing off-chip transactions.
L2H has the benefits of the classic memory hierarchy such as software transparency, simplicity of design, and isolation, while increasing  cache utilization.

%Cache underutilization is prevalent because the main consumers of server processors are public clouds, where core utilization still remains low in spite of numerous efforts~\cite{heracles,resourceCentral,memoryHarvesting,serverlessonHarvested,smartharvest,SLOHarvested}. 
%A new study conducted by Microsoft on Azure shows that 90\% of VMs run at 40\% or lower core utilization \cite{resourceCentral}. A similar study at Alibaba demonstrates that the average CPU utilization of the entire cluster is mostly between 20\%-50\% \cite{AlibabaUtil}. 
%The main reason is that users tend to over-allocate resources to ensure a prompt response to any sudden increase in load.
% The low core utilization leaves many large private L2 caches unused for the majority of the time. For instance, for a typical 64-core CPU with 1MB/core L2 caches, if the utilization is around 50\%, then 32MB cache capacity is wasted just in one CPU. Given that there are millions of servers running in public clouds including Azure, AWS, Alibaba, and Google, this huge underutilization translates to gigabytes of scarce on-chip cache wastage that could have otherwise been used to reduce memory access latency.

\begin{figure}[t]
  \centering
  \includegraphics[width=\columnwidth]{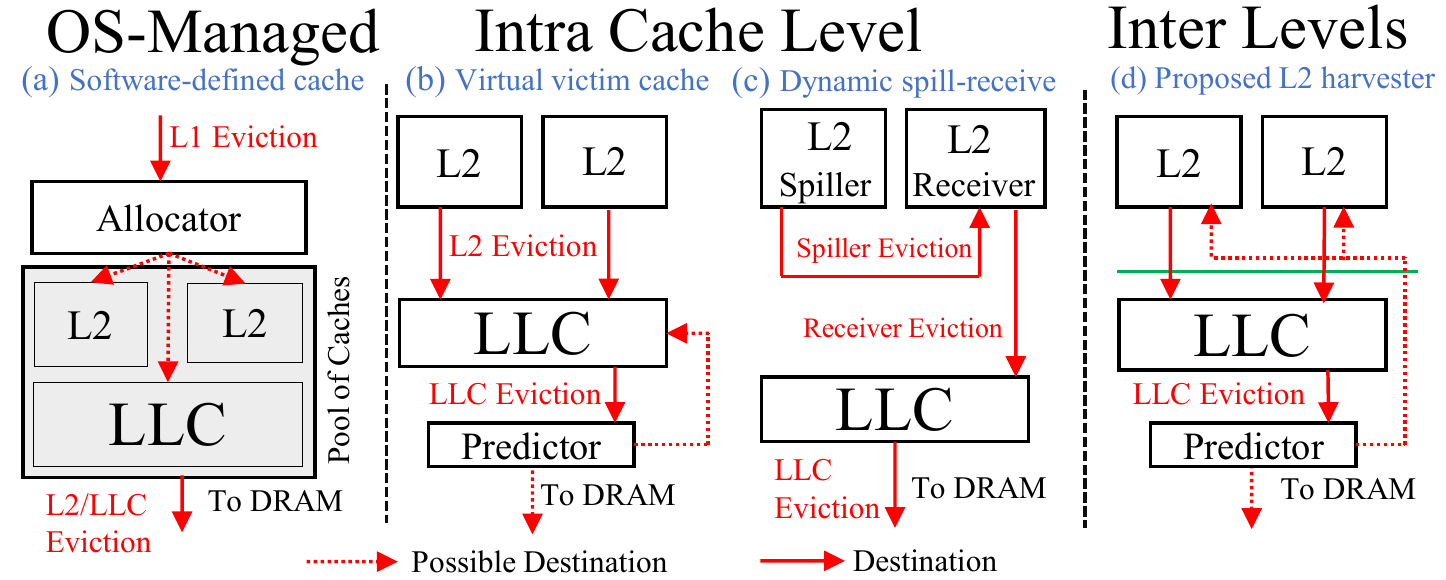}
  \caption{Comparing L2 Harvester to other solutions.}
  \label{fig:design_space_L2H}
\end{figure}

Prior work has examined different techniques to address cache underutilization. Figure \ref{fig:design_space_L2H}(a) displays OS-managed schemes, such as Jenga \cite{jenga} and IBM Z16 \cite{ibmz16}, where the cache hierarchy is flattened and an allocator determines eviction placement, adding significant complexity to design.
Figure \ref{fig:design_space_L2H}(b) presents the virtual victim cache \cite{VCache}, which aims to accommodate evictions in different sets. However, imbalancing other LLC sets exacerbates performance variation, particularly in cloud environments where predictability is critical to meeting service-level agreements.
Figure \ref{fig:design_space_L2H}(c) depicts DSR \cite{DSR} and CC \cite{CC}, which redirect evictions to other private caches rather than to LLC or DRAM. However, DSR's heuristics require transferring cache miss statistics from all caches for each eviction, resulting in a substantial increase in network traffic. Moreover, DSR does not differentiate between dead or live blocks during eviction swapping, resulting in unnecessary traffic.

% Jenga \cite{jenga} proposes a reconfigurable cache where flat, distributed, and heterogeneous cache banks are controlled and managed by hardware and run-time. Flattening the cache allows Jenga to use all cache space, but this increases the complexity of the memory sub-system. Similarly, IBM Z16 \cite{ibmz16} offers a 4-level cache hierarchy where L2 and L3 can be reconfigured to be part of an L2 cache. Z16 is shipped with Processor Resource/Systems Manager that decides how to use these large caches. In both cases, the memory hierarchy is no longer software transparent and adds huge complexity to current designs.

We revisit DSR \cite{DSR}, and address its shortcomings, including the need for transferring cache miss statistics, the blind accommodation of evictions, and extend it to a 3-level cache hierarchy. In L2 Harvester, on LLC evictions, we predict if the block is not dead and will be accessed soon. If so, the block is sent to a load balancer to decide if there is any idle core. Upon finding an available L2 cache, the block is written to the lender L2 cache, and the snoop filter metadata is updated as normal. Later, if a request to this block is received, the lender cache responds and satisfies the request. Note that L2H relies on the current coherence mechanism to locate the cache block, and does not need any special support from the hardware or runtime system, thus it is superior to designs such as Jenga \cite{jenga} and IBM Z16 \cite{ibmz16}.

L2H utilizes two lightweight predictors to find the dead blocks: (1) a bloom filter-based predictor that tracks recently evicted addresses, identifying those misses that could have been avoided with a larger cache; and (2) MPPP \cite{mp}: a perceptron-based dead block predictor that combines different features such as address and program counter to predict if a block has exhausted its useful lifetime. The two predictors complement each other: MPPP covers the bloom filter when it is not warmed up, and the bloom filter makes up for the MPPP sensitivity to thresholds when the system load is high.

We consult both predictors and make our decision based on a simple algorithm: if the load in the system is high, both predictors should agree if a block is not dead in order for the block to be sent up to a private L2 cache. If the load is low, the block is sent up if either predicts the block is not dead.

We take into account the importance and criticality of applications being run to give them a fair share of private L2 caches. User-facing applications maximally use the extra cache space as they have the highest priority in the system. Background jobs can also get extra space if the load balancer detects that the user-facing applications are not cache-sensitive, and can yield the extra space.

We evaluate L2H under different utilization scenarios. First, when the CPU load is very low ($<$25\%) and running one application. This allows the application to take up all private L2 caches in the system, representing the upper-bound benefit of L2H. Then, we move to more complex scenarios where a mix of critical and background jobs are run. L2H must make decisions regarding what blocks are dead and how to split the private L2 caches. 

We implement L2H in gem5 \cite{gem5} and run applications from different domains (datacenter, scientific, and graph analytics). Our experimental result shows that for a single application with multiple lenders, we improve P99 latency by 2$\times$. Also, for mixes of user-facing and background jobs, L2H improves P99 and throughput by up to 32\% and 50\%.
 
To summarize our main contributions:
\begin{itemize}
 \item We demonstrate that a substantial amount of cache capacity is wasted in modern processors due to a rigid hierarchal design, and conservative resource allocation in the cloud.

 \item We architect and evaluate an effective, yet low-cost L2 harvesting mechanism that enables a logical path from LLC evictions to private L2 caches. This allows the idle cores to lend their unused L2 caches, thus keeping more data blocks on the chip.

 \item We incorporate two dead block prediction schemes in the L2 harvester to identify those capacity/conflict-caused evictions that are worth keeping on chip. We also devise a simple load balancer that distributes data blocks over unused resources by taking system load and criticality of applications into the account.

\item We evaluate our proposed method and compare it to a conventional hierarchy with a larger LLC. Our evaluation results show that a quad-core system with 2MB/core LLC and 1.25 MB/core L2 cache benefiting from L2H improves system performance by up to 2$\times$ over the baseline. Also, we show that L2H provides competitive system performance compared to a baseline with a 50\% larger LLC (12MB).

\end{itemize}

\section{Motivation}
\label{sec:motivation}

\begin{table}[]
  \begin{adjustwidth}{-7pt}{}
    \scriptsize
    \caption{Intel and AMD CPU generations.}
    \label{tab:gens}
    \begin{tabular}{ll|l|l|l|l|l}
      & \multicolumn{3}{c|}{\intel Intel (2016-2020)}                             & \multicolumn{3}{c}{\amd AMD (2017-2022)}  \\
                                    & \intel SKX               & \intel CSX                & \intel ICX               & \amd Rome            & \amd Milan(X)             & \amd Genoa          \\ \cline{2-7}
      \multicolumn{1}{l||}{L1}      & \intel 32KB              & \intel 32KB               & \intel 48KB              & \amd 32KB             & \amd 32KB             & \amd 32KB              \\
      \multicolumn{1}{l||}{L2/core} & \intel 1MB               & \intel 1MB                & \intel 1.25MB            & \amd 512KB            & \amd 512KB            & \amd 1MB             \\
      \multicolumn{1}{l||}{L3/core} & \intel 1.37MB            & \intel 1.37MB             & \intel 1.5MB             & \amd 4-8MB            & \amd 4-12MB            & \amd 4-16MB            \\
      \multicolumn{1}{l||}{Cores}   & \intel 4-28              & \intel 2-56               & \intel 8-40              & \amd 8-64             & \amd 8-64             & \amd 8-96              \\ \hline
      \multicolumn{1}{l||}{Total}   & \intel 66.5MB            & \intel 133MB              & \intel 110MB             & \amd 288MB             & \amd 800MB            & \amd 1100MB             \\
      \multicolumn{1}{l||}{SKU }    & \intel \tiny{Xeon-P8180} & \intel  \tiny{Xeon-P9282} & \intel \tiny{Xeon-P8380} & \amd \tiny{EPYC-7H12} & \amd \tiny{EPYC-7773X} & \amd \tiny{N/A}
    \end{tabular}
  \end{adjustwidth}
\end{table}

According to Microsoft Azure and Alibaba, datacenter core utilization is very low. Servers run at 40\% or lower utilization at 90\% of the time at Azure \cite{resourceCentral}, and between 20\%-50\% most of the time at Alibaba \cite{AlibabaUtil}. This over-allocation stems from the fact that VMs should have enough cores and resources if the load surges rapidly.

In addition, CPU manufacturers are increasing the L2/LLC sizes and the number of cores. \tab{tab:gens} exhibits three generations of Intel and AMD server CPUs. We can observe that both manufacturers' L2/LLC and core counts have steadily increased over generations. L2 and LLC sizes are reaching 1.25MB/core, 2MB/core for Intel processors, and 1MB/core and 4MB/core for AMD processors. Combined with the fact that core counts are also increasing, we can see that the third generation of Intel processors are accumulating 110MB total cache capacity, while AMD is reaching over giga bytes of on-chip cache storage.

\begin{figure}[t]
  \centering
  \includegraphics[width=\columnwidth]{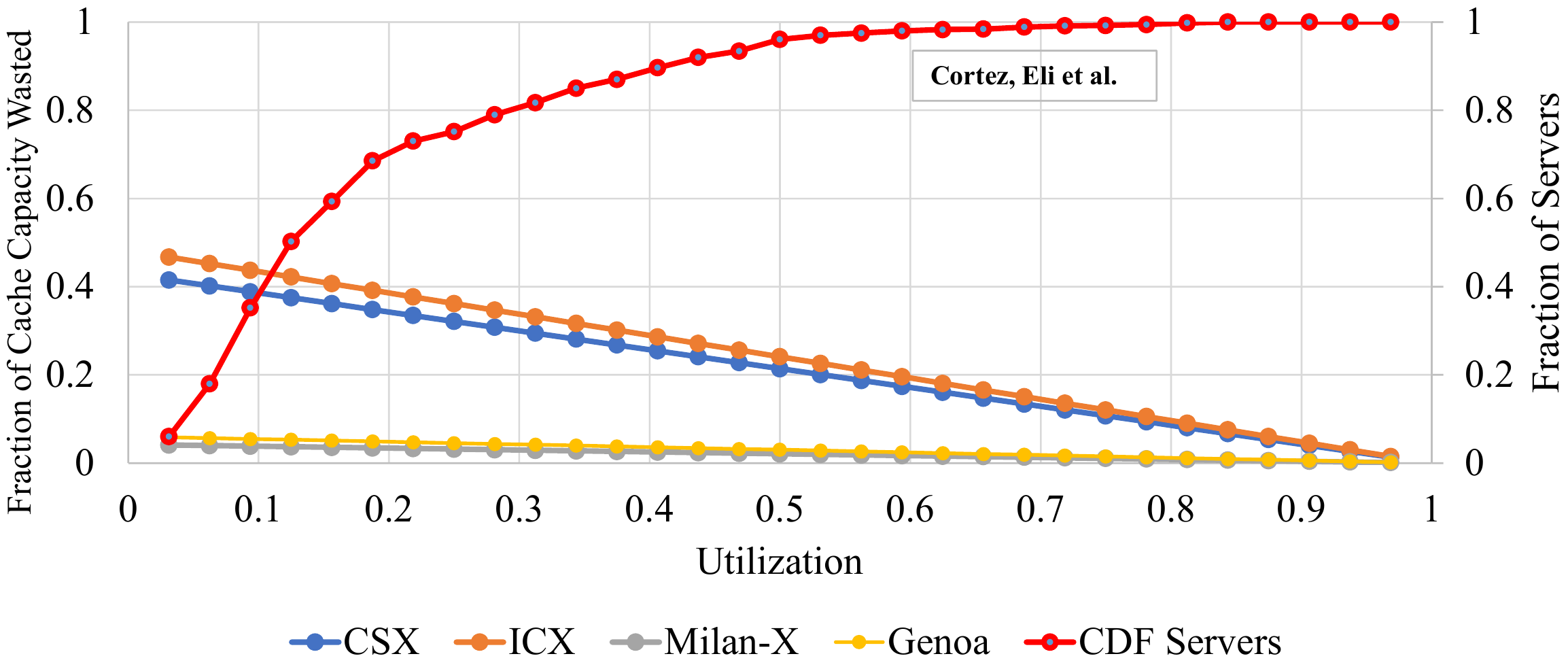}
  \caption{Total cache wasted by different server processors under various utilization. The CDF (red line) is taken from \cite{resourceCentral}.}
  \label{fig:harvestOpp}
\end{figure}

To better understand the current situation in datacenters, \fig{fig:harvestOpp} shows the total cache capacity wasted by different server processors under various utilization levels.
On the x-axis, we show the utilization. We assume that all processors have 32 cores. Thus, the minimum utilization is when there is one application running taking one core and the whole LLC ($\frac{1}{32}=0.03$), and maximum utilization is when all 32 cores are active ($\frac{32}{32}=1$).
To calculate the total cache wasted (first y-axis), we subtract the used cache capacity under each load from the total cache capacity available on the chip ($32~\times (L1 + L2)+LLC$). For example, if there are two cores running, and L1=48KB, L2=1MB/core, and LLC=8MB, then wasted cache is $32\times(48KB+1MB)+8MB-2\times(48KB+1MB)-8MB$.
On the second y-axis, we show the CDF of core utilization on Azure \cite{resourceCentral}.

\begin{table}[t]
\caption{Total L2 cache capacity of 3 supercomputers in the TACC datacenter.}
\label{tab:tacc}
\begin{tabular}{lcccc}
Systems   & Nodes & Processor            & Core/node & Total L2 (GB) \\ \hline
Frontera~\cite{frontera}  & 8008  & Xeon 8280 & 56        & 438                  \\
Lonestar6~\cite{ls6} & 560   & EPYC 7763      & 128       & 35                   \\
Chamealon~\cite{ch} & 10000 &  Haswell       & 96        & 469                 
\end{tabular}
\end{table}

\begin{figure*}[t]
  \centering
  \includegraphics[width=2.2in]{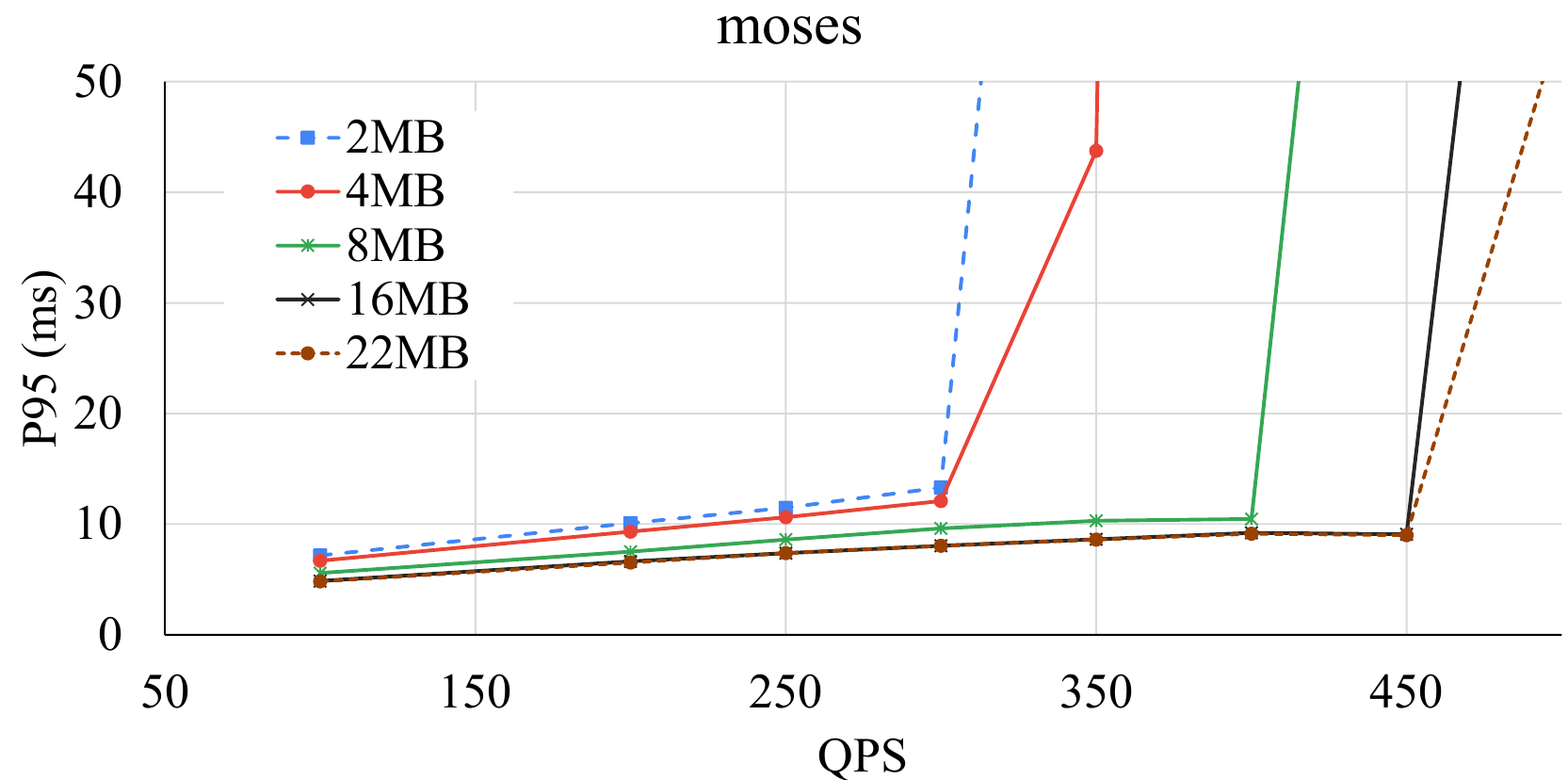}
  \includegraphics[width=2.2in]{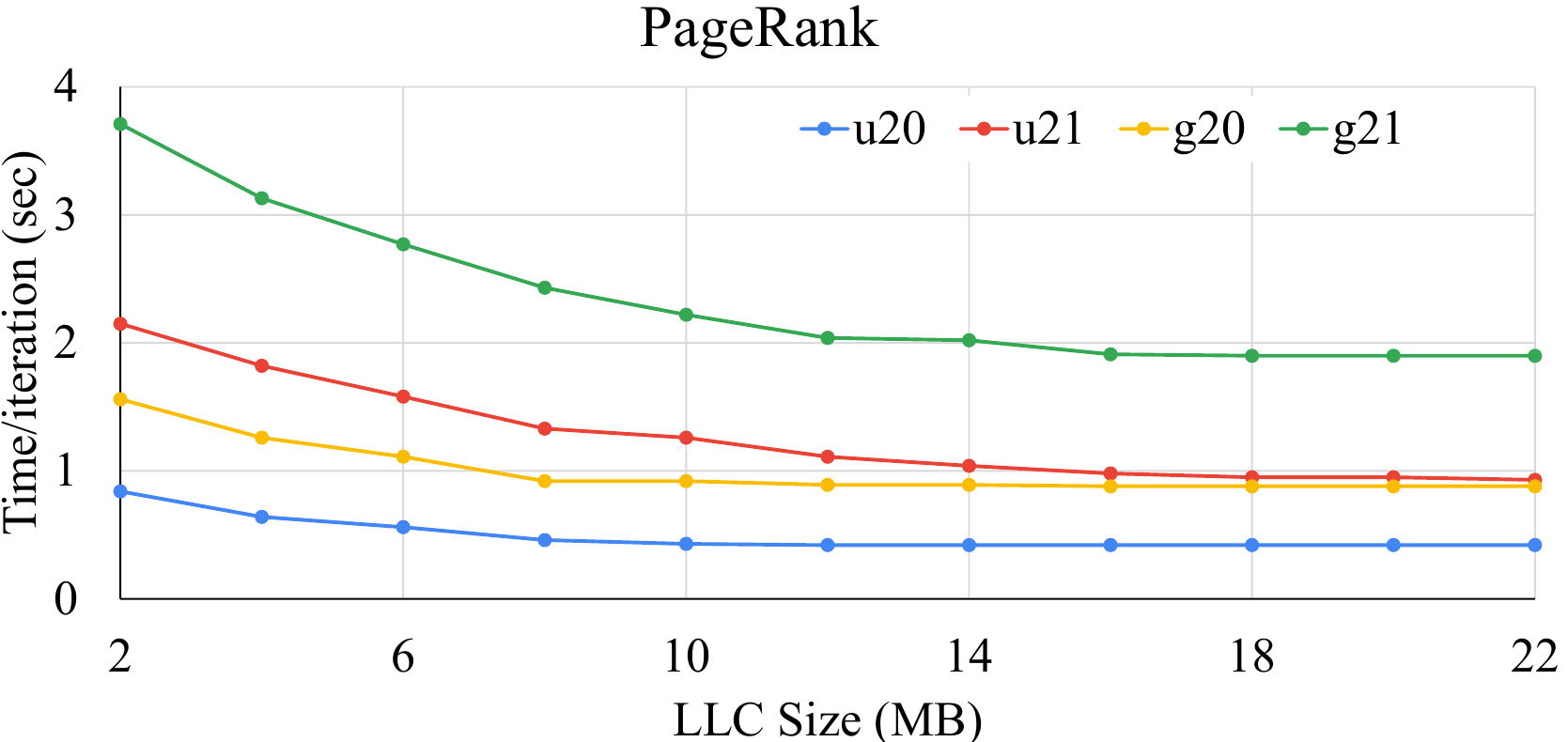}
  \includegraphics[width=2.2in]{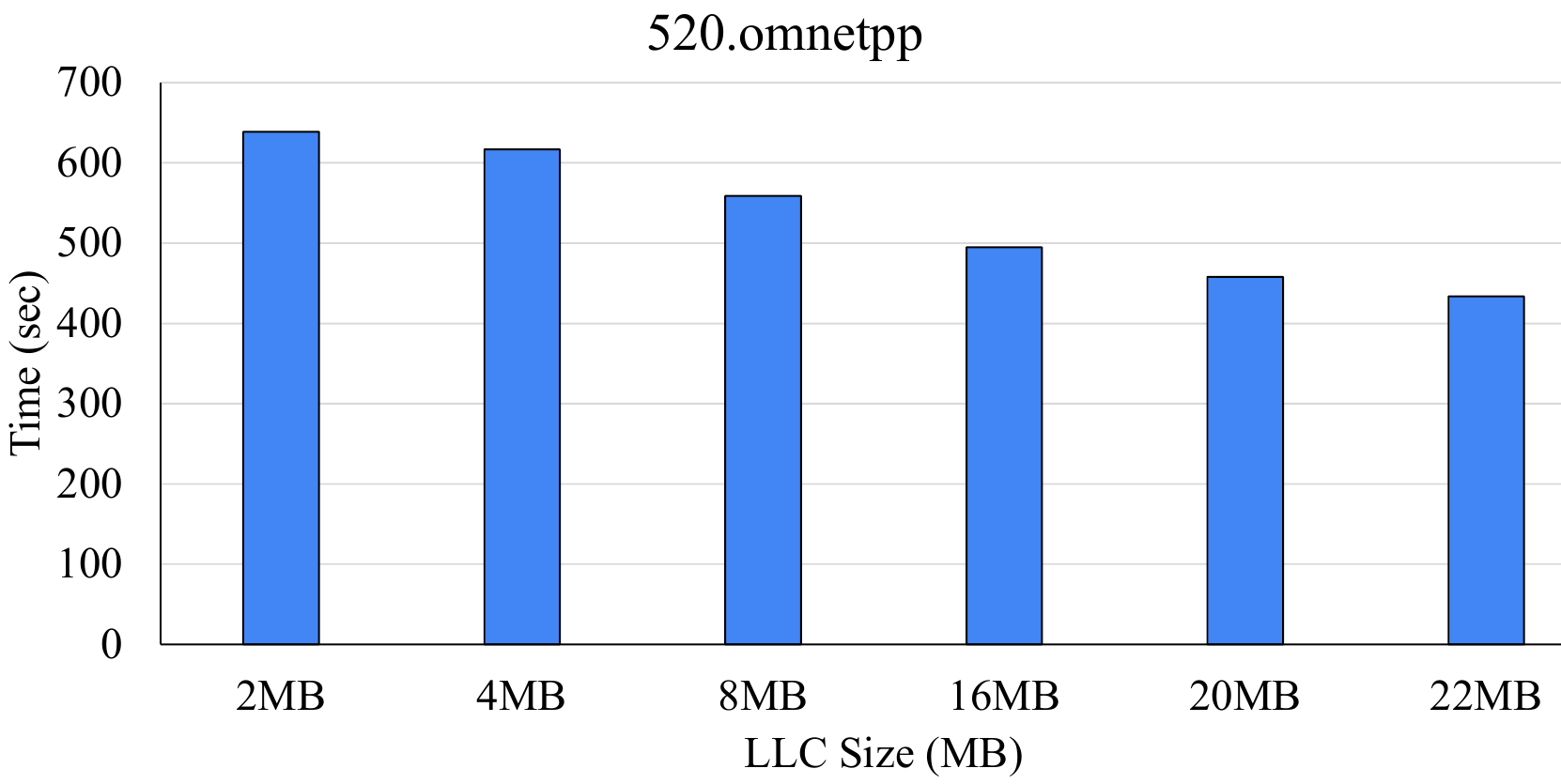}
  \caption{Impact of LLC size on applications performance.}
  \label{fig:mot_real}
\end{figure*}

As can be seen from \fig{fig:harvestOpp}, 50\% of Azure Icelake machines waste around 40\% of the total cache capacity. Given that Icelake machines have an L2 capacity of 1.25MB/core (see \tab{tab:gens}), for 32 cores, around 35MB of total on-chip cache capacity is wasted that could otherwise be used to keep data blocks on the chip and boost up system performance.
AMD processors also suffer from similar issues but at smaller scale. For instance, Rome wastes around 10\% of cache capacity under the load of 40\%. The main reason is that AMD has smaller L2/core, and very large L3/core capacity. However, in Genoa, we observe that AMD is enlarging the L2/core from 512KB/core in Milan to 1MB/core.

To put L2 cache waste into perspective, \tab{tab:tacc} shows 3 supercomputers in the TACC datacenter (Frontera, Lonestart6, and Chameleon). The table shows the main processor types as well as the number of nodes and total L2 capacity (GB). As can be seen, the total cache capacity in a small-scale datacenter like TACC can be somewhere between 35GB (Lonestart6) to 469GB (Chameleon). Hence, if the utilization is around 50\% on average, a substantial amount of a very scarce resource like L2 cache is being wasted (234.5GB in Chameleon and 17.5GB in Lonestart6). Note that public clouds such as AWS, Azure, Google, and Alibaba are operating significantly larger datacenters, so we are projecting the L2 waste reaches to terabytes. 

A larger cache capacity can help reduce the long memory access latency. We conduct a cache study on a real machine to measure how much cache capacity impacts system performance. The machine is an Intel(R) Xeon(R) Gold 6242 CPU with 22MB 11-way LLC cache. We run one application and change the cache size using Intel Cache Allocation Technology (CAT) from one way (2MB) to 11 ways (22MB). We set the core frequency to 3.9GHz.

\fig{fig:mot_real} shows the performnace for three applications: (1) \emph{moses} from TailBench \cite{TailBench}, where we sweep the system load in terms of query per second (qps) and cache size; (2) \emph{PageRank} from gapbs \cite{gapbs} with 4 different synthetic inputs (u: uniform graph, and g: Kronecker graph), and two different sizes (20, and 21); and (3) \emph{520.omnetpp} from SPEC CPU 2017 \cite{SPEC17}.

For \emph{moses} we make two observations: (1) with larger caches, the saturation point (point that P95 increases sharply) is pushed to higher qps (further to the right). For example, we can see that the knee point for 22MB occurs at 450, while for the 2M, the server is saturated at qps=300; 1.5$\times$ improvement in the maximum load; (2) at similar loads before the saturation point the larger caches provides better P95 latency. For instance, when qps=250, we see that 2MB LLC provides P95 of 12ms, while the 22MB cache shows P95 of 8ms.

For \emph{PageRank} we observe that a larger LLC reduces the execution time significantly. For example, for the largest graph  (g21) the execution time is halved when increasing the LLC size from 2MB to 12MB. We also see that for LLC sizes of greater than 12MB, the execution times remain fixed. 
Finally, for \emph{520.omnetpp} we observe similar sensitivity to cache size. the execution time constantly reduces from 610 seconds for 2MB LLC, to 420 seconds for 22MB cache. \emph{Our conclusion is that larger cache help applications from different domains, thus wasting a huge amount of on-chip cache is not reasonable, and we need to devise schemes to allow the unused L2 caches to be utilized when possible.}

%%% Local Variables:
%%% mode: latex
%%% TeX-master: "main"
%%% End:

\section{L2 Harvester $\mu$architecture}
\label{sec:proposed}
We propose L2H, a simple yet effective mechanism for harvesting L2 caches, that provides performance improvement for memory-bound applications. In this section, we first overview the design of L2H. Then, we discuss the algorithm behind detecting the dead blocks, and how we distribute the blocks over idle cores.

\subsection{L2H: Overview and Organization}
\begin{figure}[t]
\centering
\includegraphics[width=\columnwidth]{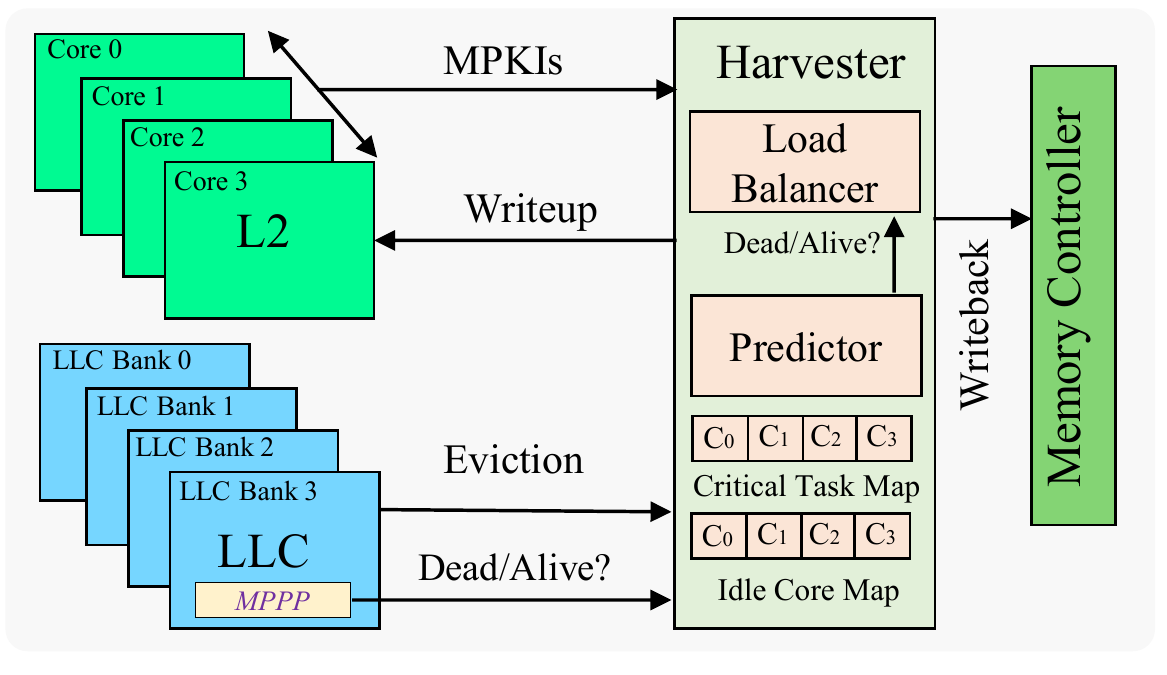}
\caption{L2 harvester architecture. MPPP \cite{mp} is the state-of-the-art dead block predictor.}
\label{fig:arch}
\end{figure}

\fig{fig:arch} shows the overview of L2H. Without loss of generality, we assume there are 4 cores connected to LLC banks with a shared bus. LLC has MPPP dead block predictor \cite{mp}. L2H sits between LLC and the memory controller and tracks the writebacks.
If a block is detected by the predictor to be not dead, is sent to the load balancer. Then, the load balancer decides where this block can be written to. If there is any idle core that can lend its L2 cache, the load balancer pushes the block up to the lender. Otherwise, if the block is dead, or if there is no free L2, the block is written back to main memory. Thus, in the next reference to this block, there might be a private L2 cache that responds to the request and thereby saves one off-chip transfer.

L2H needs four pieces of information to perform prediction and load balancing:
(1) L2 MPKIs;
(2) Critical Task Map (CTM): a bit mask that determines if the application being run on a core is critical, ``1'' determines the application being run at core \emph{n} is critical. This bit mask is provided by the user or system administrator and is updated as soon as a new application is assigned to cores;
(3) Idle Core Map (ICM): a bit mask that determines if a core is idle and can lend its L2 cache. This is updated by the cores if core has nothing to execute; and
(4) The output of the MPPP \cite{mp} dead block predictor.

\subsection{L2H Structures}
\noindent\textbf{Predictor} The purpose of the predictor is to determine if a block is dead, and thus it is not worth keeping on chip. This is particularly important for streaming applications because redirecting all cache blocks to upper levels will waste power, increase traffic, and elevate congestion on coherence.

\begin{figure}[t]
\centering
\includegraphics[width=2.6in]{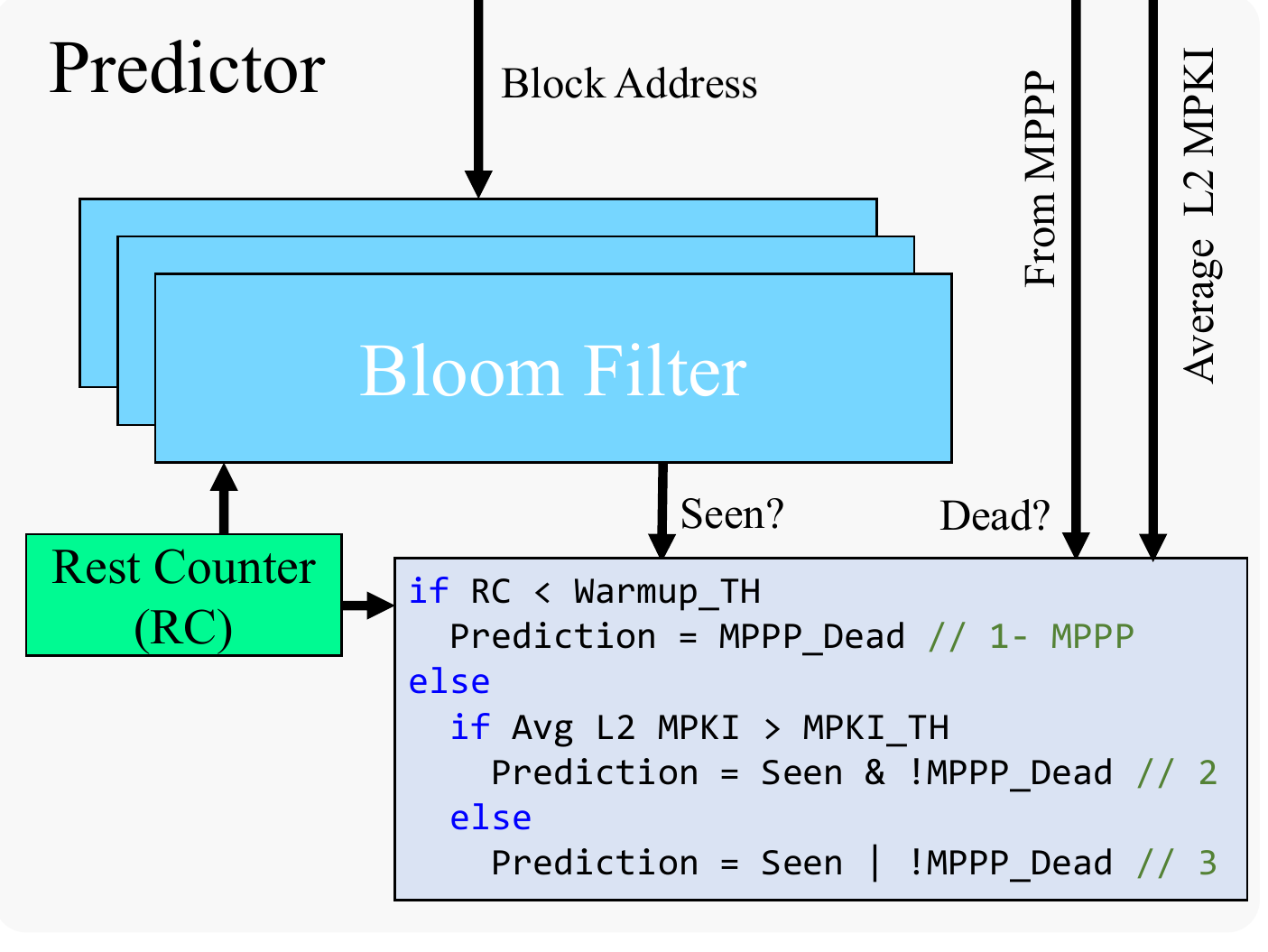}
\caption{L2 harvester predictor.}
\label{fig:predictor}
\end{figure}

\fig{fig:predictor} shows the structure of our predictor. We combine two predictors to find dead blocks: (1) a bloom filter-based predictor; and (2) the multi-perspective perceptron predictor (MPPP) \cite{mp}.
The functionality of the bloom filter-based predictor is simple. We insert the missed addresses into the bloom filter. To make a prediction, we just need to look up the address, if the address was not found in the filter, we conclude the block is dead. Because we have \emph{not} seen a reference to this block recently. We reset the bloom filter periodically to make sure the false positive rate stays low. Unfortunately, after each reset, the bloom filter starts declaring all blocks dead as they have not been seen, thus we need to address this shortcoming.

Morpheus \cite{morpheus} uses a bloom filter for hit/miss prediction, and addresses this problem by using two separate bloom filters with different reset intervals. So, when one of them is being warmed up, the other one services the requests, and vice versa. However, we found that we get better accuracy if we combine our bloom filter with another type of dead block predictor (e.g., perceptron-based dead block predictor). The two predictors complement each other: MPPP covers the bloom filter when it is not warmed up, and the bloom filter makes up for the MPPP sensitivity to thresholds when the system load is high.

We use MPPP to solve the reset problem of the bloom filter. MPPP \cite{mp} is a perceptron-based technique that predicts the future reuse of cache blocks. MPPP combines several features including program counter and address to form weight tables. Then taking summations of entries from each table, it predicts if a block is: (a) not dead, (b) dead on arrival, and can bypass the cache, and (c) dead, and can be evicted from the cache. MPPP uses three thresholds to make the prediction based on the aggregated values taken from the weight tables.

Our experiments show that MPPP works well when MPKI in the system is not very high, but it becomes very sensitive to the thresholds when MPKI is very high. The issue is that when there are many misses, the MPPP tables are updated more frequently; we increase the value for one entry and decrement for the rest (usually cache associativity -1). This lead to a situation where MPPP observes smaller aggregated values. Hence, differentiating dead blocks becomes more challenging. However, this is a situation where the bloom filter works well, because it warms up faster, and can help to detect the addresses that have been evicted recently.

Hence, while the bloom filter is being warmed up, we use MPPP to find the dead blocks, and we rely on the bloom filter when the load is high and MPPP becomes sensitive to the thresholds. The second advantage is that for challenging applications, we can refer to both predictors to decide if a block is dead to increase the accuracy.

As can be seen from \fig{fig:predictor}, when a block arrives, and if the bloom filter is \emph{not} warmed up ($RC<Warmup\_TH$), then we have no options other than relying on MPPP for prediction ($Prediction=MPPP\_Dead$). Otherwise (if the bloom is warmed up $RC>Warmup\_TH$), then we have both predictors available to make a prediction. In such a case, if the load is high (L2 MPKI $> MPKI_{TH}$) both predictors should agree on the outcome ($Prediction = Seen~\&~!MPPP\_Dead$). Otherwise, the block is \emph{not} dead, if either predictor predicts so ($Prediction = Seen~|~!MPPP\_Dead$).

\begin{figure}[t]
\centering
\includegraphics[width=3in]{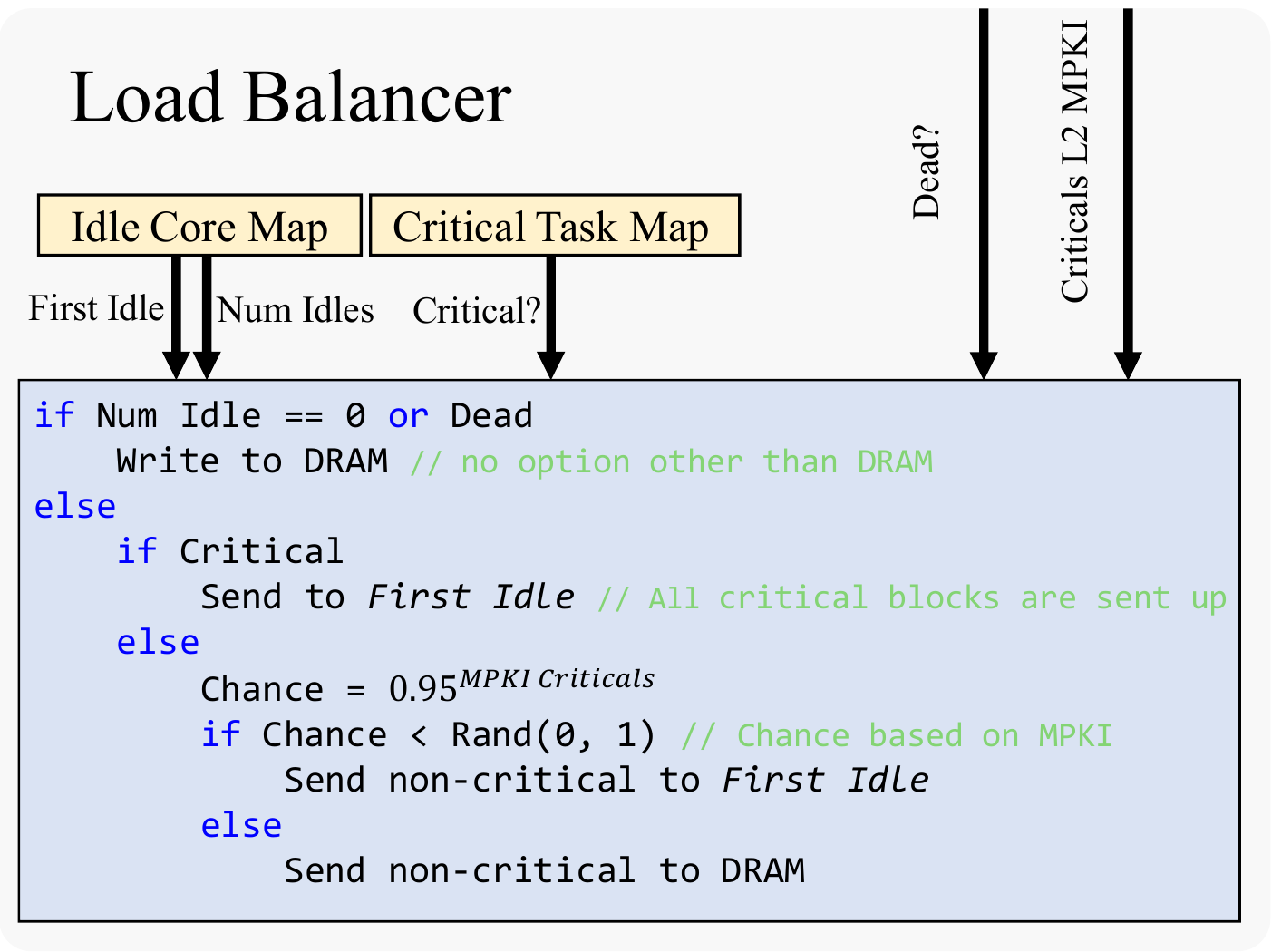}
\caption{The overview of the load balancer.}
\label{fig:load_balancer}
\end{figure}

\noindent\textbf{Load Balancer} The purpose of the load balancer is two-fold: (1) find a lender and make decision if a block must be sent up; (2) redirect dead blocks, and non-critical live blocks to the main memory if the system load is high. \fig{fig:load_balancer} shows the structure and algorithm of the load balancer.

The load balancer takes as the input five pieces of information:
(1) output of the predictor as a boolean signal called $Dead$;
(2) average L2 MPKI of caches running user-facing applications;
(3) first idle L2 cache obtained from Idle Core Map (ICM) using a round-robin scheme;
(4) a boolean signal named $Critical$ if this block belongs to the core running critical applications; and
(5) total number of critical applications running at the moment in the system obtained from Critical Task Map (CTM).

The intuition behind the load balancer algorithm is to give critical applications with maximum L2 capacity and provide the non-critical applications with as much as the capacity that will not negatively impact the critical applications. The algorithm works as follows: if there is no idle core, or if the block is dead, we must write the block back to main memory. If there is an idle core, and if the block belongs to critical applications, it will be pushed to the first idle resources.

\begin{figure}[t]
\centering
\includegraphics[width=2.8in]{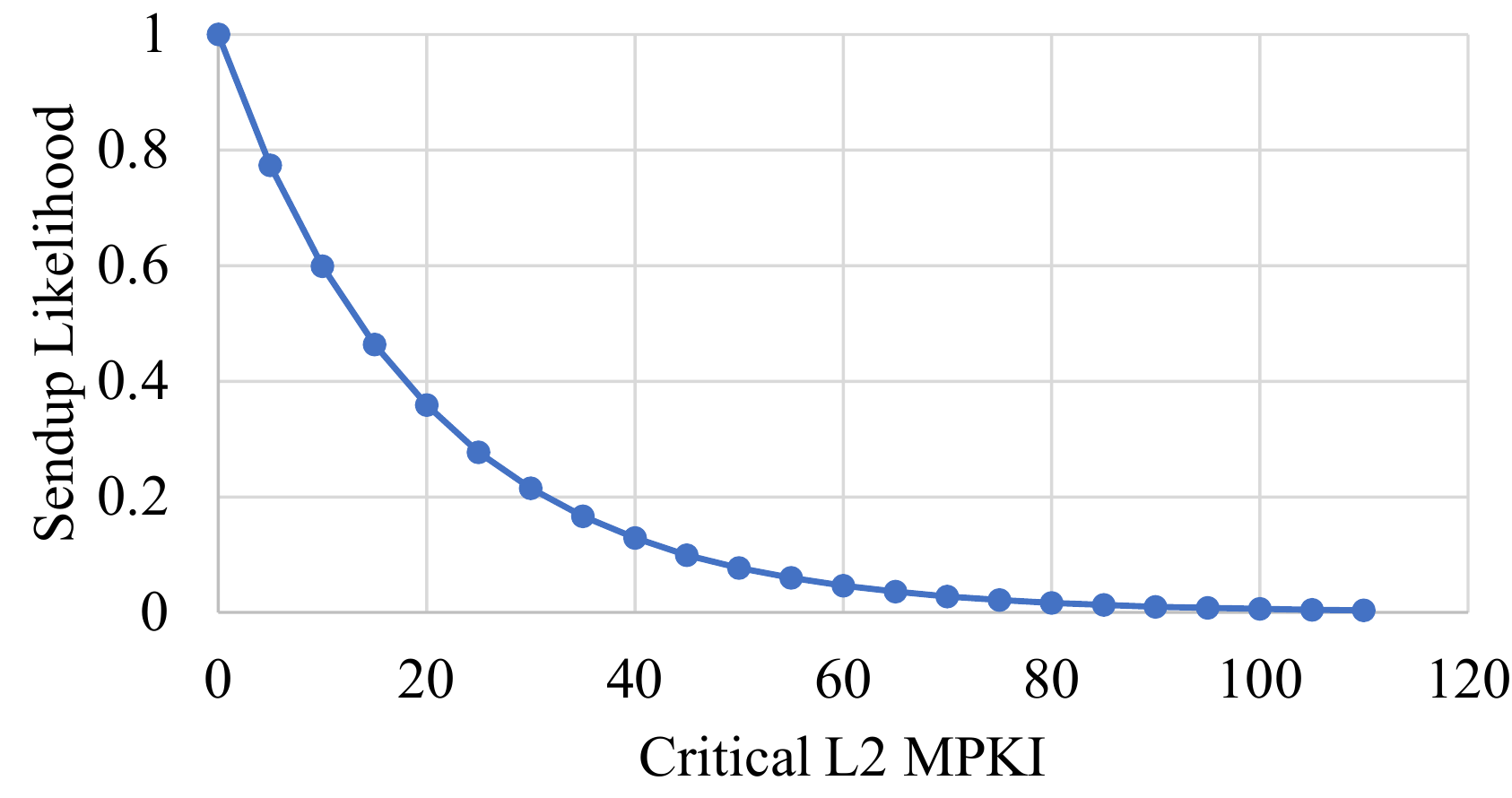}
\caption{Probability of sending a non-critical block to a private L2 cache as a function of critical applications MPKI.}
\label{fig:wchance}
\end{figure}

On the other hand, if the block is not critical, we probabilistically send the block to a private L2 cache with a probability that decays as critical L2 MPKI grows. The intuition is that requests should not be sent up when L2 MPKI is high. We arbitrarily choose an exponentially decaying probability density function ($Chance=0.95^{MPKI}$) as shown in \fig{fig:wchance}. Hence, if the MPKI is low for critical applications, we give a fraction of the capacity to the non-critical applications. As the MPKI for critical applications increases, the chance for non-critical applications decreases. For example, if the MPKI=20, the chance of sending a non-critical application reduces to 30\%, while for $MPKIs>40$, non-critical blocks will be barely sent to the private caches.

\subsection{L2 Harvester Operation}
\begin{figure}[t]
\centering
\includegraphics[width=\columnwidth]{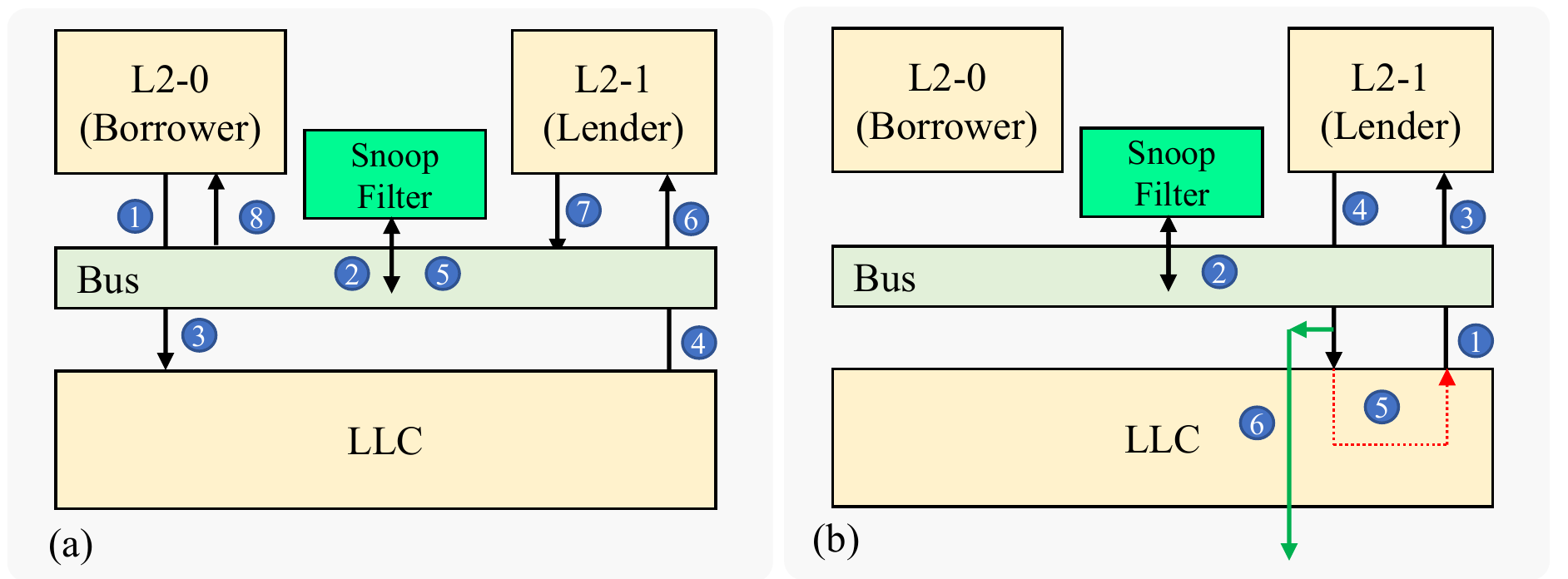}
\caption{(a) L2H operations; and (b) Circular problem.}
\label{fig:operation}
\end{figure}

\fig{fig:operation} shows how the harvester works in practice. As \fig{fig:operation}(a) shows, in Step \ballnumber{1} a cache block is evicted from its private L2 cache, sends over the bus and checks the snoop filter in Step \ballnumber{2} to find its destination port. The snoop filter directs the block to the LLC. This block stays in the LLC until it is evicted in Step \ballnumber{4}. The L2 harvester decides to send it to L2-1. The block lookups the snoop filter in Step \ballnumber{5}, updates its location to be L2-1, and is filled in the lender in Step \ballnumber{6}. Later, when a request to this block arrives, the snoop filter redirects the request to the lender (L2-1), and the response is sent back by the lender to the borrower in Step \ballnumber{8}. Note that we do not change the functionality of the snoop filter; this operation is treated as a normal transfer to L2-1.

\noindent\textbf{Possible Circular Harvesting} L2H may create a circular situation where a block stays on the chip and never gets evicted despite not being useful. \fig{fig:operation}(b) shows such a scenario. Similar to the previous example, assume that in Step \ballnumber{1} a block is redirected to a lender. Thus, it updates the snoop filter (Step \ballnumber{2}) and fills in the cache (Step \ballnumber{3}). Eventually, this block gets evicted and is sent to the LLC in Step \ballnumber{5}. Upon eviction from the LLC, it may again be redirected to a private L2 cache based on a prediction. This loop can happen infinitely, and this cache block will never depart the chip, even though it is not touched. To address this problem, we add one extra bit to the L2 cache tag store indicating if a block has been redirected to the upper-level cache. Then, when we are evicting this cache from the private cache, instead of writing it back to the LLC, we bypass the LLC in Step \ballnumber{6} and write it directly to the main memory. We find this approach to help because this block has been given a second chance already and can be evicted from the cache to avoid creating circular harvesting.

\section{Evaluation Methodology}
\label{sec:eval}
\subsection{Simulator Configuration}
We use the gem5 full-system cycle-level simulator to conduct the experiments~\cite{gem5}. We model a 3-level cache hierarchy where L1 and L2 are inclusive and private and L3 (the LLC) is non-inclusive and shared. L1, L2, and L3 are parallel caches where tag and data stores are accessed in parallel. L1 is 12-way 48KB/core with 1-cycle access latency, L2 is 12-way 1.25MB/core with 12-cycle access latency, and the LLC is 16-way 2MB/core at 25-cycle access latency. We use one prefetcher per level: L1 uses AMPM \cite{AMPM}, L2 runs DCPT \cite{DCPT}, and LLC uses STeMS \cite{STeMS}. L1, L2, and the LLC have 16, 32, 64 MSHR entries. 

We also find that always enabling these prefetchers significantly degrades system performance for some applications (e.g., 505.mcf) because the prefetchers contend too strongly with demand requests. We, therefore, implement two prefetch throttling mechanisms. In the first scheme, we reserve 25\% of MSHR entries for demand accesses, which decreases the prefetch rate and maintains some minimum demand request service. The second throttling mechanism is that we monitor the performance of the prefetcher periodically and disable a prefetcher when its accuracy drops below 40\%. Specifically, in each epoch of 10 million accesses, the prefetchers operate for the first 1 million accesses, then the prefetcher accuracy determines if the prefetcher remains enabled for the following 9 million accesses.

We use the MPPP \cite{mp} dead block predictor for the LLC. Similarly to the original design, we use all 16 features suggested by the authors for multicores as listed in \tab{tab:features}. We use 256 randomly selected sets to train the model. When a block is accessed in the cache, all features are extracted from the address, and program counters and used to index the weight tables. Then, we sum up all weights and if it exceeds a threshold, the block is declared dead. To train the model, when any of the sampled sets are accessed (fills or hits in the sampler as suggested in the paper), we extract the features from the access. Then, we use the features to look up the tables and increase the counters. Also, we decrement the counters associated for those blocks that are impacted by sampled access's promotion. 

MPPP \cite{mp} does not explicitly provide the thresholds in the paper. Hence, in order to find the threshold to declare a block dead, we ran 10 experiments, each running 4 randomly chosen applications and swept the thresholds comparing the MPPP suggestions with those of the bloom filter. We found that if the summation of features is greater than 320, MPPP exhibits the best performance. We refer the reader to MPPP~\cite{mp} for more detail.

For the bloom filter, we use the structure proposed by Sanchez et al.~\cite{bf_h3}. The bloom filter has 4096 entries and 4 hash functions. This bloom filter uses a high-quality hash functions (H3~\cite{H3}). Given that redirecting evictions is not on a critical path, we do not use parallel bloom filter lookup, and instead use a single-port structure to save power and area. 

\begin{table}[]
\label{tab:features}
\caption{Features used to form MPPP tables~\cite{mp}: feature(LRU stack position, start bit, end bit, [$n^{th} access$], [XOR with PC]).}
\begin{tabular}{|llll|}
\hline
bias(6,0)           & addr(9,9,14,5,1) & addr(9,12,29,0) & addr(13,21,29,0) \\
addr(14,17,25,0) & lastmiss(6,0)       & lastmiss(18,0)     & offset(13,0,4,0)    \\
offset(14,0,6,0)    & offset(16,0,1,0)     & pc(6,13,31,4,0)    & pc(9,11,7,16,0)     \\
pc(13,16,24,17,0)   & pc(16,2,10,2,0)     & pc(16,4,46,9,0)    & pc(17,0,13,5,0)   \\ \hline
\end{tabular}\end{table}

The main memory is DDR4-3200. There is one command and address bus, with timings based on a DDR4-3200 8Gbit device (Micron MT40A1G8) in an 8 × 8 configuration. The total channel capacity is 16GB. This maintains a reasonable core-to-memory ratio for the simulations.

The core has 320, 128, and 128 ROB, LQ, and SQ entries, respectively. The core frequency is set to 3.66GHz. Fetch-, commit-, and writeback-widths are all set to 8. The branch predictor is TAGE\_SC\_L \cite{tage}. The TLB has 128 entries, and there are 8 page-table walkers. 

\subsection{Benchmarks}
We evaluate the applications of: (1) Tailbench \cite{TailBench} representing user-facing jobs in datacenters; (2) SPEC CPU 2017 \cite{SPEC17} representing background jobs; and (3) gapbs graph analytics benchmarks.
We mainly choose applications that are memory-bound and benefit from larger cache capacity, but also include some compute-bound applications to show how the proposed solution behaves in such scenarios. 

We choose 2 memory-bound applications from Tailbnech (\emph{moses} and \emph{img-dnn}) and one compute-bound application (\emph{massstree}). \emph{moses} is a statistical machine translation (SMT) system. The input is randomly-chosen dialogue snippets from the \textit{opensubtitles.org} English-Spanish corpus. \emph{moses} has high L2 and LLC MPKIs of 26, and 22, respectively. \emph{img-dnn} is a handwriting recognition that uses OpenCV under the hood. The input to this application is chosen randomly from MNIST dataset. \emph{img-dnn} shows L2 and LLC MPKIs of 20, and 18, respectively. We also evaluate \emph{masstree} fast key-value store applications written in C++. This application has MPKIs of 6 and 5, respectively. \emph{masstree} is driven with the Yahoo Cloud Serving Benchmark.

We choose 5 memory-bound applications from SPEC CPU 2017: \emph{502.gcc},  \emph{505.mcf}, \emph{519.lbm}, \emph{520.omnetpp}, and \emph{549.fotonik3d}. We also run 3 compute-bound applications: \emph{500.perlbench}, \emph{531.deepsjeng}, and \emph{521.wrf}. From gapbs, we choose 3 applications: the page rank algorithm to find the web page ranking (\emph{pr}), the betweenness centrality score for approximate calculations all vertices in a graph by only computing the shortest paths from a subset of the vertices (\emph{bc}); and single-source shortest paths that computes the distances of the shortest paths from a given source vertex to every other reachable vertex (\emph{sssp}).

We drive \emph{pr}, \emph{bc}, and \emph{sssp} with synthetic graphs: (1) \emph{u}: a synthetically generated graph by the Erddos–Reyni model (Uniform Random); and (2) \emph{g:} a synthetically generated graph by the Kronecker synthetic graph generator. We set the input size to be $2^{20}$ and $2^{21}$. Note that all applications of gapbs are memory-bound, and thus we do not have any compute-bound representative application from this suite.

\subsection{Single-Application Runs}
We run \emph{moses}, \emph{masstree}, and \emph{img-dnn} for 250 requests on gem5: We launch Tailbnech in integrated mode, where both client and server are running within one process. Then, we warm up the internal data structures by running 1000 requests in fast-simulation mode via KVM CPUs. After the warm-up is finished, we switch the simulator CPU model to the most accurate version (detailed OOO), and continue the simulation until 250 requests are serviced. Due to the fact that clients and the server are run in one process, architectural statistics are not accurate. Hence, we record request timestamps while the applications are running on top of the simulator, and copy them back to the host, and calculate the P99 of simulated 250 requests. 

For SPEC CPU, we use the SimPoint methodology \cite{simpoint} to find representative regions of each application. We use 2 SimPoints of 250 million instructions each and 250 million instructions for warmup. 
For gapbs, we run each application 10 times after the graph was generated.

\subsection{Multi-Applications Runs}
We use Tailbench to represent the user-facing latency-critical applications, and SPEC CPU 17 and gapbs applications as background tasks. Due to gem5 limitations, simulating more than 4 cores is very slow and difficult. Hence, we limit our study to 4 cores. For user-facing applications we choose one application from img-dnn, masstree, and moses, and one application from SPEC CPU 2017, or gapbs. We leave two cores idle each can provide 1.25MB L2 cache. Similar to the single-application scenario, we run the user-facing applications for 250 requests and make sure the background job continues to run until the simulation is finished. We create 50 random mixes out of the applications listed in \tab{tab:mix}.

\begin{table}[t]
\scriptsize
\renewcommand{\arraystretch}{1.2}
\caption{Evaluated System Configuration.}
\label{Table:System}
\centering
\resizebox{\linewidth}{!}{%
\begin{tabular}{|p{0.6in}||p{2.3in}|}
\hline
Processor & Single and Quad-core, 3.66 GHz, Ubuntu 20.04 OS. ROB:320, LQ:128, SQ:128, Fetch-width=8\\
\hline
L1 Cache & 48kB 8-way; 12 ways; LRU; 1 cycles. Prefetcher: AMPM \cite{AMPM}\\
\hline
L2 Cache & 1.25MB 12-way; LRU; 16 ways; 12 cycles,  Prefetcher: DCPT\cite{DCPT} \\
\hline
L3 Cache & 2MB/core; 16-way; LRU; 25 cycles. Prefetcher: STeMS \cite{STeMS}\\
\hline
Main Memory & 16~GB:  DDR4-3200 x64, 8x8 Micron MT40A1G8\\
\hline
\end{tabular} }
\end{table}

\begin{table}[t]
\caption{Multi-program applications.}
\label{tab:mix}
\begin{tabular}{|l|ll|}
\hline
User-facing & \begin{tabular}[c]{@{}l@{}}img-dnn\\ qps=200, 300, 400\end{tabular} & \begin{tabular}[c]{@{}l@{}}masstree\\ qps=200, 300, 500\end{tabular} \\ \hline
Background  & \multicolumn{2}{l|}{bc\_u20, pr\_u20, sssp\_u20, sjeng, omnet, lbm, mcf, perl} \\ \hline                                                            
\end{tabular}
\end{table}

\subsection{Systems}
We compare three 4-core systems: (1) baseline with an 8MB LLC; (2) the baseline configuration but with a 12MB LLC; and (3) L2H with an 8MB LLC. Depending on the number of applications running, L2H can borrow 3, 2, or 1 L2 caches. Hence, the total L2 and L3 capacity for L2H is 8MB LLC + 3$\times$1.25MB=11.75MB at most when it borrows three L2 caches, and 8MB LLC+ 1$\times$1.25MB=9.25MB, when it borrows one L2 cache.

\section{Evaluation Results}
\label{sec:results}
%---------------------------
\begin{figure*}[t]
\centering
\includegraphics[width=2.2in]{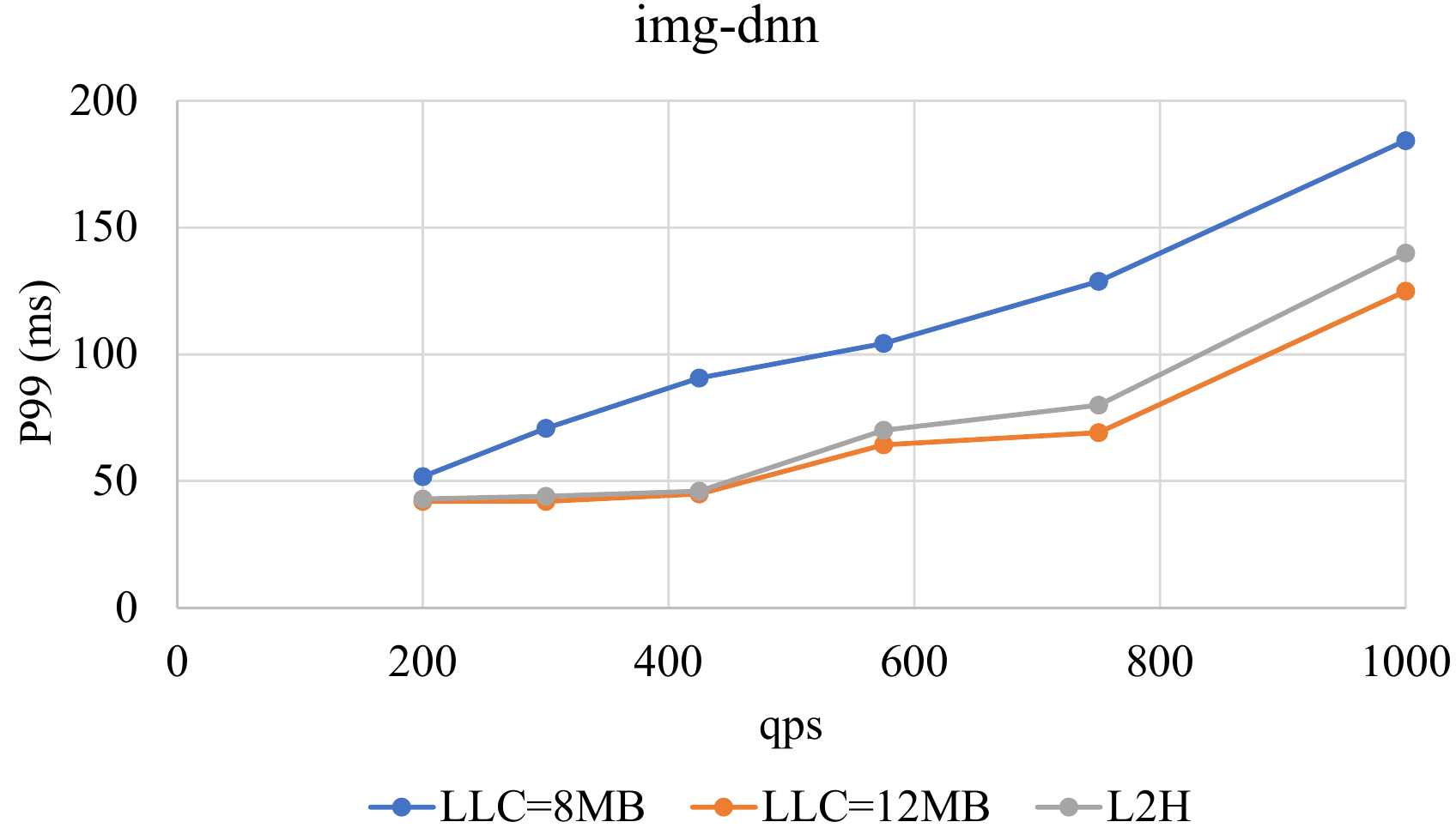}
\includegraphics[width=2.2in]{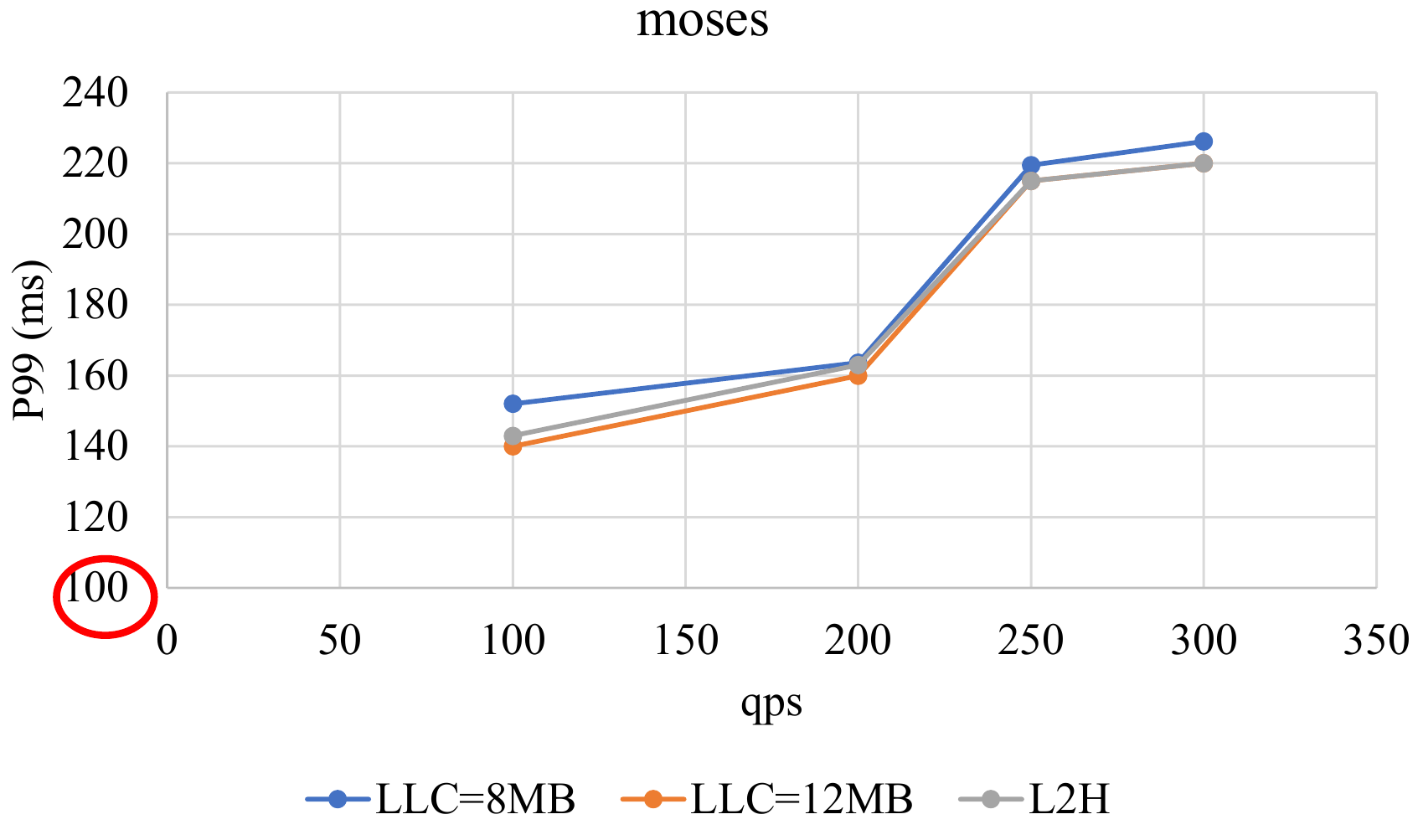}
\includegraphics[width=2.2in]{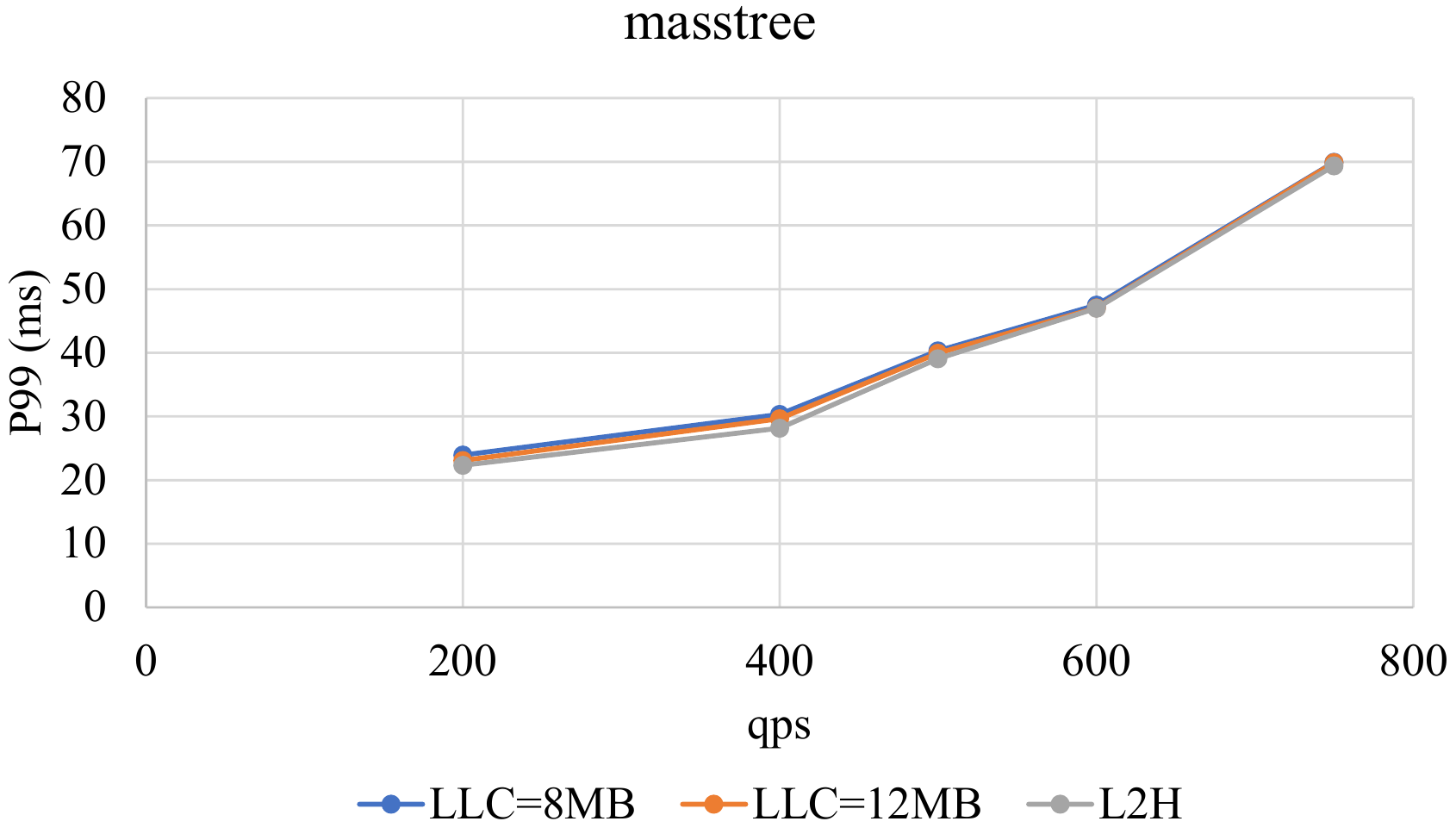}
\caption{Impact of LLC size, and QPS on applications performance for (1) baseline with 8MB LLC; (2) baseline with 12MB LLC; and (3) L2H with 8MB LLC, and 3 lenders each 1.25MB.}
\label{fig:IPC_single_tail}
\end{figure*}
%------------------------
%------------------------
\begin{figure*}[t]
\centering
\includegraphics[width=4.4in]{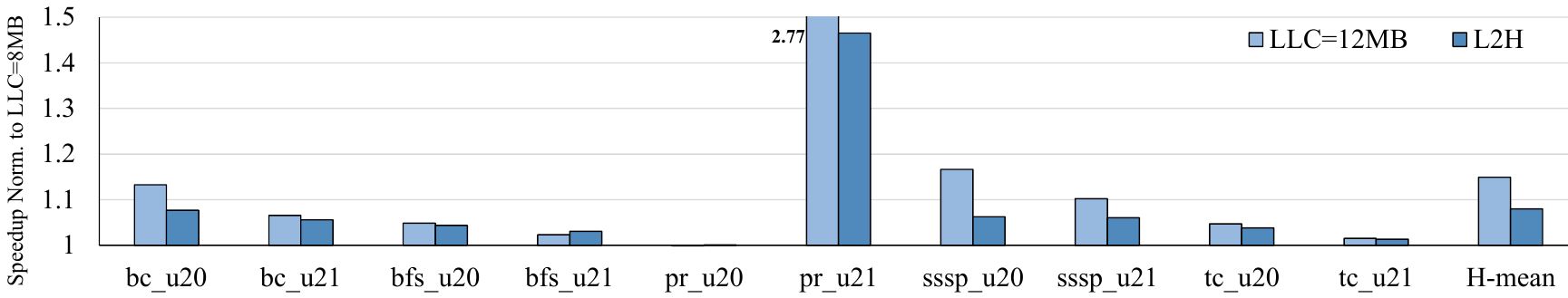}
\includegraphics[width=2.3in]{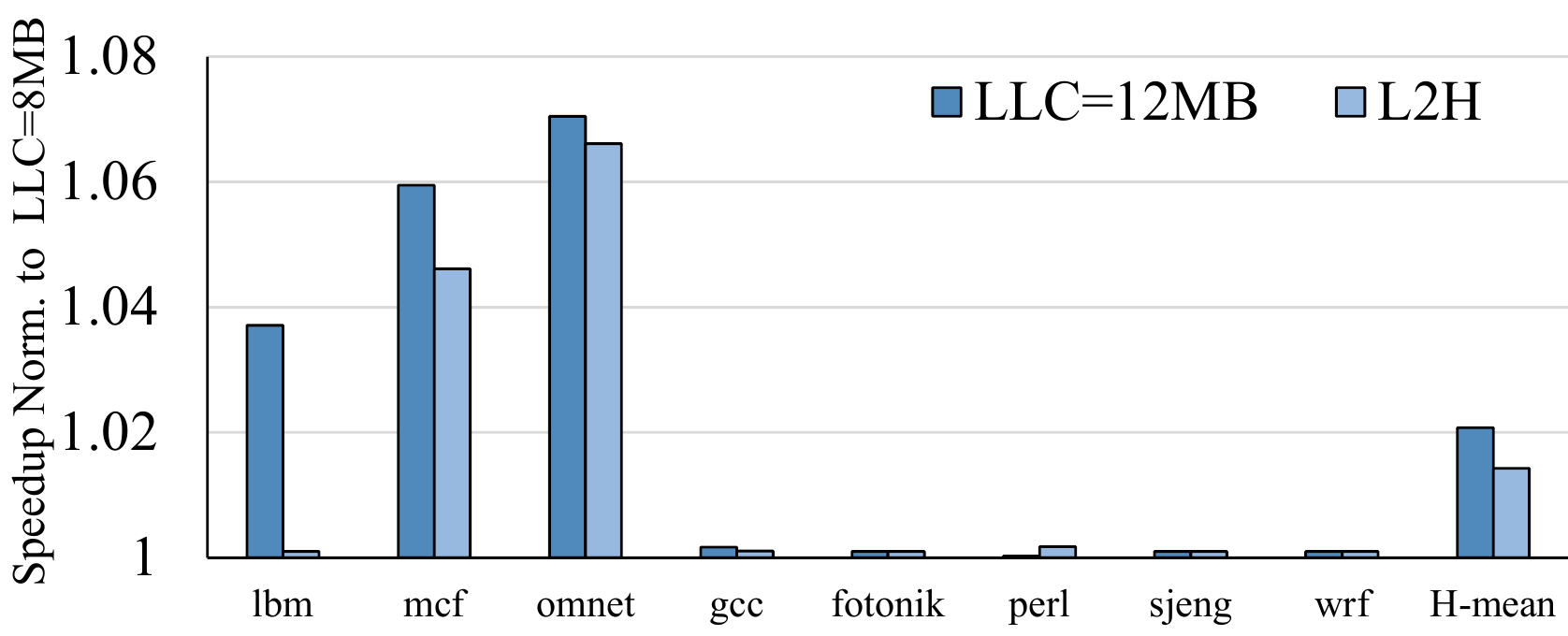}
\caption{Impact of LLC size on applications performance for (1) baseline with 8MB LLC; (2) baseline with 12MB LLC; and (3) L2H with 8MB LLC, and 3 lenders each 1.25MB.}
\label{fig:IPC_single_spec}
\end{figure*}
%------------------------

\subsection{Single Application with Three Lenders}
\noindent\textbf{Performance} In this section, we analyze a scenario where one application is running, and there are three idle cores (25\% utilization) lending their private L2 caches.
\fig{fig:IPC_single_tail} shows the impact of LLC configuration on the latency-throughput curves in terms of P99 latency. We compare three LLC configurations (8MB, 12MB, or 8MB+L2H with 3.75MB of borrowed L2 capacity) on three user-facing applications (\emph{img-dnn}, \emph{moses}, and \emph{masstree}). 

\begin{figure*}[!htb]
    \begin{minipage}{4.5in}
        \includegraphics[height=1.0in,left]{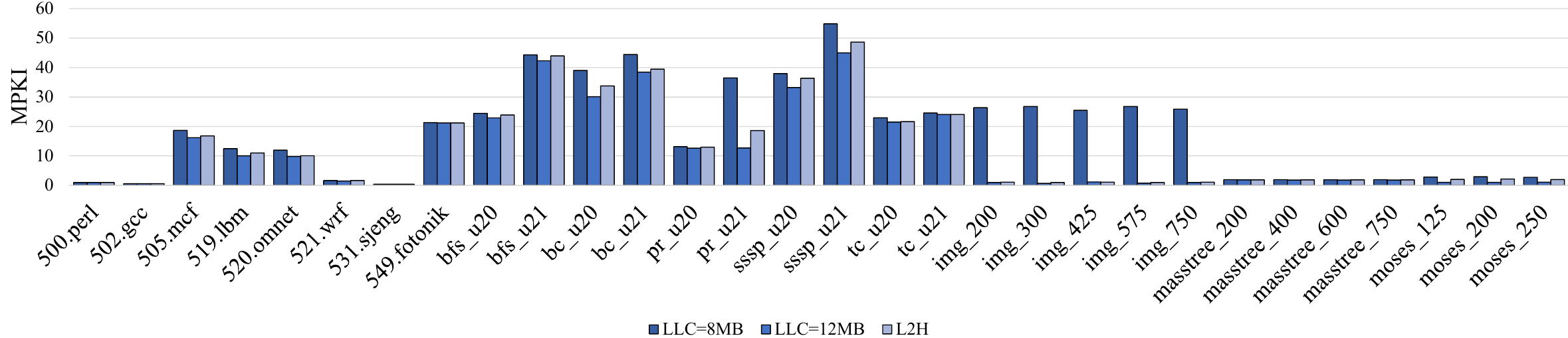}
        \caption{MPKI for 3 systems: (1) an 8MB LLC; (2) a 12 MB LLC; and (3) L2H.}
        \label{fig:single_mpki}
    \end{minipage}%
    \begin{minipage}{2.2in}
        \centering
        \includegraphics[height=.9in,left]{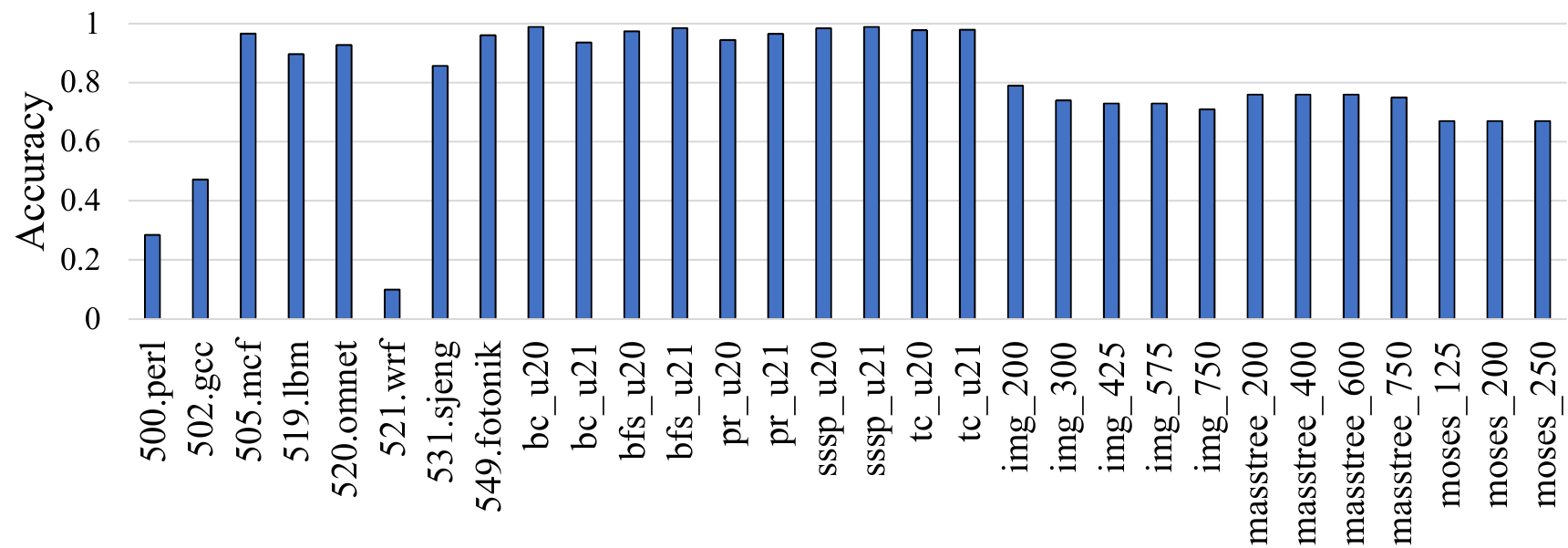}
        \caption{Prediction accuracy. Fraction of blocks that the load balancer sends to private L2 caches that satisfy a request.}
        \label{fig:single_acc}
    \end{minipage}
\end{figure*}

\emph{img-dnn} benefits from the larger cache the most. L2H closely follows the 12MB LLC, while the gap between these two and the 8MB LLC stays fairly constant (2X better P99). The reason for such a large performance improvement can stems from the large reduction in MPKI. As shown in \fig{fig:single_mpki}, the \emph{img-dnn} MPKI decreases from 26 to 2 when the LLC size reaches 12MB. This implies that \emph{img-dnn} working set size fits in the larger LLC, and thus a huge P99 improvement is realized. \emph{L2H could provide the needed capacity for such applications almost for free with a 33\% smaller LLC size (8MB vs.\ 12MB).}

\emph{moses} performance is shown in \fig{fig:IPC_single_tail}(middle). At the lowest qps (100), \emph{moses} shows 7\% and 5\% lower P99 for a 12MB LLC and L2H compared to the baseline with 8MB LLC, respectively. L2H closely tracks the 12MB LLC. %The gap between L2H and LLC=12MB slightly bridges as the qps increases, but we see that L2H stays beneficial (3\% better than LLC=8MB at qps=300).

\fig{fig:IPC_single_tail}(right) shows the performance of \emph{masstree}, whose MPKI is very low (1.1). This application is not memory bound, so we do not expect to see improvement in P99 when the LLC grows. We also expect L2H to not negatively impact the P99 latency. As expected, all three systems show very similar P99 latency, meaning L2H does not interfere with compute-bound applications. We observe similar behavior (not shown) across other compue-bound applications as well (\emph{shore}, \emph{xapian}, \emph{specJBB}, and \emph{silo}).

In addition to lowering the P99 latency, extra cache space can increase the maximum supported load: the qps after which the P99 latency increases sharply.  For \emph{img-dnn} the saturation point is pushed to higher qps by the 12 MB LCC and L2H: baseline with 8MB LLC has a rapid increase in P99 for qps$>$200, but the saturation point occurs at qps=500 for both L2H and the 12MB LLC.

\fig{fig:IPC_single_spec} shows system performance on the gapbs and SPEC CPU 2017 benchmark suites. The harmonic mean speedups for L2H are 15\% and 1.7\% for gapbs, and SPECU CPU 2017, respectively. Among gapbs application, page rank with the \emph{u:21} input exhibits the largest speedup (2.77$\times$ for the 12MB LLC and 1.26 $\times$ for L2H). As with \emph{img-dnn} case, the MPKI of  \emph{pr\_u21} decreases from 36 to 15. 

SPEC CPU applications also benefit from larger caches, but to a lesser extent. We found that only 3 applications somewhat benefit from larger caches in this benchmark suite: \emph{omnet} 6.2\%, \emph{505.mcf} 4.5\%, and \emph{lbm} 3.9\%. However, the majority of applications do not significantly benefit from the larger caches. We found two reasons for this behavior: (1) some applications are cache-friendly, but an 8MB LLC is sufficient for them; and (2) other applications such as \emph{perl} and \emph{wrf} are not memory-bound, and their MPKIs are less than 2.

%----------------------

\noindent\textbf{MPKI} \fig{fig:single_mpki} shows the MPKI for the three LLC configurations. The normalized geo-mean performance of the 12MB LLC and L2H are 15\% and 12\% better than the baseline with an 8MB LLC, respectively. Note that L2H achieves this 12\% better MPKI with 33\% less LLC size (8MB vs.\ 12MB). This brings a substantial saving in terms of area, power and cost.  We make two observations: (1) there are applications such as \emph{pr} and \emph{img-dnn} whose MPKIs are reduced significantly due to fitting the whole working set in the cache; and (2) there are applications with various MPKI ranging from 1 to 55 in our evaluation, stressing the load balancer properly.

\noindent\textbf{Prediction Accuracy} \fig{fig:single_acc} shows the prediction accuracy of L2H. We calculate the accuracy by counting how many blocks are sent up and what fraction of those are requested by the borrower. The average prediction accuracy for memory-bound applications is 89\%. The averages are 96\%, 75\%, and 70\% for gapbs (applications with the highest MPKIs), Tailbench, and SPEC CPU 2017, respectively. There are some applications with low prediction accuracy such as \emph{perl}, \emph{gcc}, and \emph{wrf}, but given that their MPKIs are very low ($<2$), the mispredictions have insignificant impact.

%-------------------
\begin{figure*}[t]
  \centering
  \includegraphics[width=\textwidth]{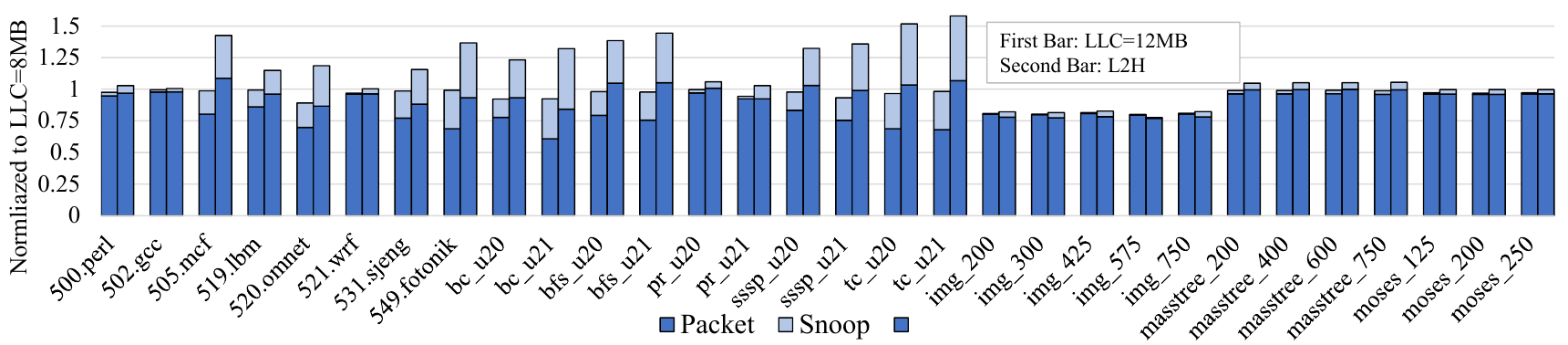}
  \caption{Traffic increase on the bus between L2s and LLC normalized to a baseline with 8MB LLC. The first bar is a baseline with 12MB, and the second bar is L2H. The dark color is the data packets, and the light blue is the snoop packets. }
  \label{fig:single_traffic}
\end{figure*}
%-------------------
%-------------------
\begin{figure}[t]
  \centering
  \includegraphics[width=\columnwidth]{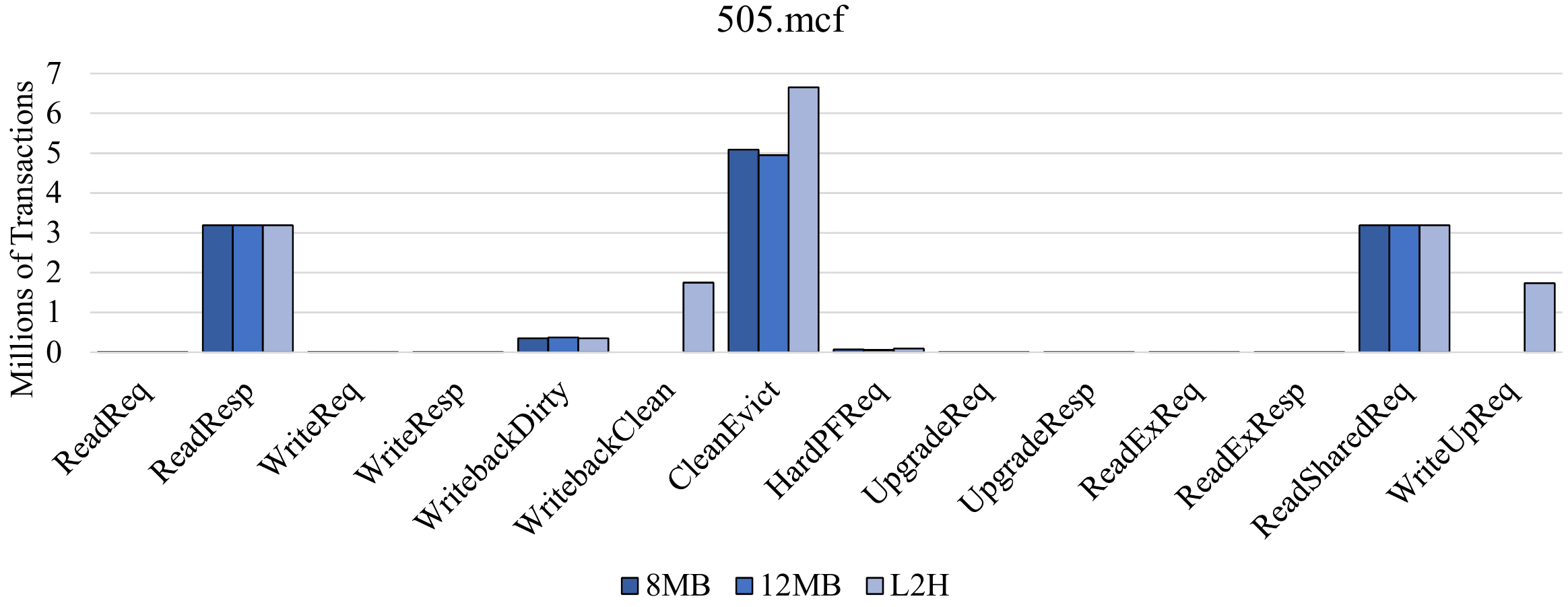}
  \caption{Example of high prediction accuracy (96\%), and high traffic (42\%). Breakdown of packets seen on the bus between L2s and LLC for \emph{505.mcf}. L2H is the only one that has \emph{WriteUp} packets. }
  \label{fig:single_mcf_traffic}
\end{figure}
%-------------------

\noindent\textbf{Traffic Analysis} L2H sends data blocks to upper-level caches based on a heuristic. Although the prediction accuracy is high, we need to carefully study any increased traffic on the shared interconnect. \fig{fig:single_traffic} shows the traffic for the 12MB cache and L2H normalized to the baseline traffic of the 8MB LLC: the First bar is the 12MB LLC and the second bar is L2H. We also separate the actual packets from the snoop packets, as they usually have different sizes and purposes: the dark blue represents actual packets and the light blue represents the snoop packet seen on the interconnect connecting L2 caches to LLC.

As can be seen from \fig{fig:single_traffic}, the 12MB LLC has consistently lower or equal traffic. This is expected because the larger cache keeps more data blocks on the chip than the 8MB LLC, so it does not generate more traffic. In terms of packet count, we can see that the majority of packets are data packets and not snoop, as there is only one application running. Given that we are running in full-system mode, the OS processes are running on the cores and may share data blocks, but this is negligible. Hence, overall, the 12MB LLC has less traffic.

\begin{figure}[t]
  \centering
  \includegraphics[width=\columnwidth]{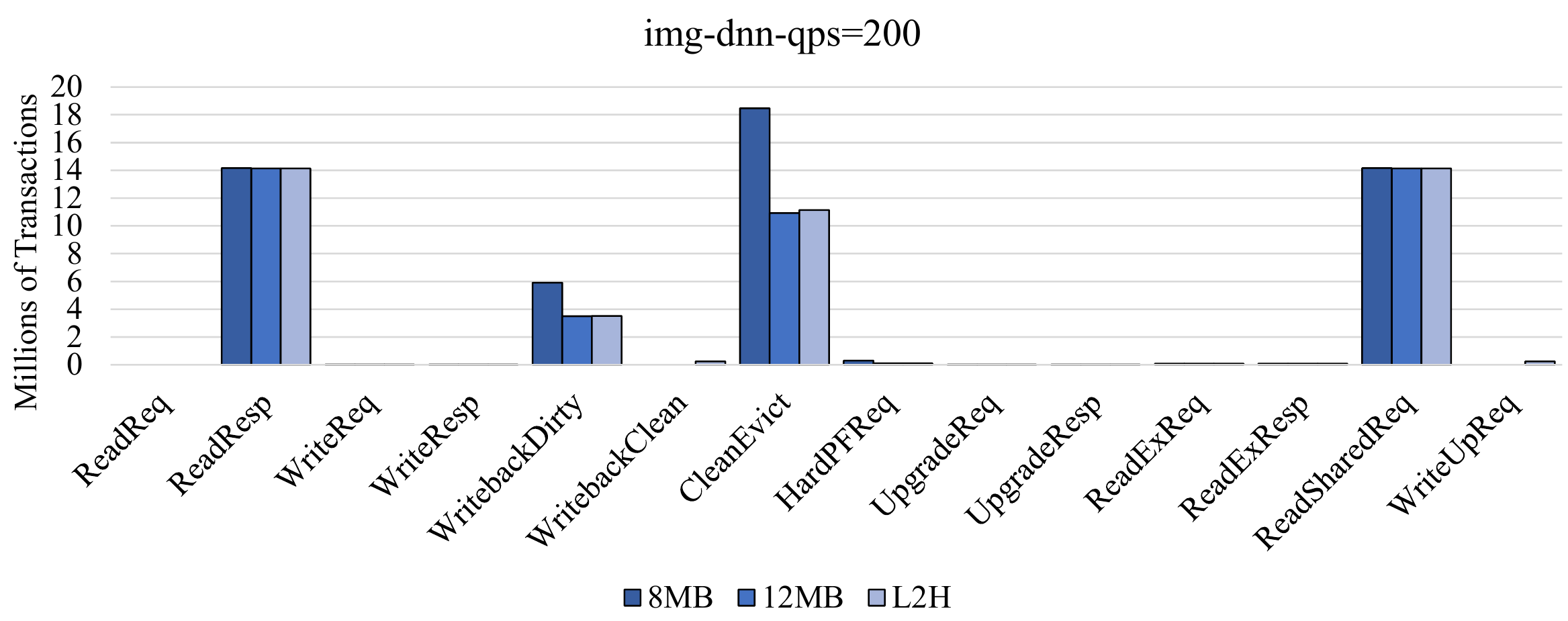}
  \caption{Example of mediocre prediction accuracy (75\%), and low bus traffic (-20\%). Breakdown of packets seen on the bus between L2s and LLC for \emph{img-dnn}. L2H is the only one that has \emph{WriteUp} packets.}
  \label{fig:single_img_traffic}
\end{figure}

On the other hand, the geo-mean for L2H is 24\% increase in traffic. This increase in traffic is expected as the blocks are sent up and distributed over private L2 caches. However, the behavior of L2H is very dynamic: some applications, such as \emph{mcf} generate more traffic (42\% more), while others, like \emph{img-dnn} generate less traffic (-20\%). Compute-bound applications (those applications for which L2H has no impact) exhibit no change in traffic. To understand this behavior better, we show the breakdown of packets for two applications in \fig{fig:single_mcf_traffic} and \fig{fig:single_img_traffic}.

The increase in traffic comes from two sources: (1) sending blocks up to a private cache, indicated as \emph{WriteUp} requests in \fig{fig:single_mcf_traffic} and \fig{fig:single_img_traffic}; (2) evicting a block that has been sent to a private cache (without first reusing it). The load balancer and the predictor accuracy determine how many \emph{WriteUpRequest} are generated. Given that prediction accuracy is high in L2H, we believe that extra traffic generated by \emph{WriteUps} will actually help performance.

L2H increases snoop traffic because it first checks the snoop filter before sending up a block. This ensures that data is not needlessly replicated.  Depending on the data block status (clean or writeback clean), this snoop request is either CleanEvict or WritebackClean. This snoop check is the main reason why we see an increase in \emph{WritebackClean} and \emph{CleanEvict} in \fig{fig:single_mcf_traffic} and \fig{fig:single_img_traffic}.

We observe that for \emph{mcf} (prediction accuracy=96\%, traffic increase=42\%), \emph{WritbackClean}, and \emph{CleanEvict} are substantially higher than the baselines, leading to a situation where the total traffic increases by 42\%. On the other hand, for \emph{img-dnn} because the larger cache can fit the working set size, the \emph{CleanEvict} for L2H stays very close to that of the 12MB LLC, helping to reduce the total traffic by 20\%.

\subsection{Two Applications with Two Lenders}
\begin{figure}[t]
  \centering
  \includegraphics[width=3.2in]{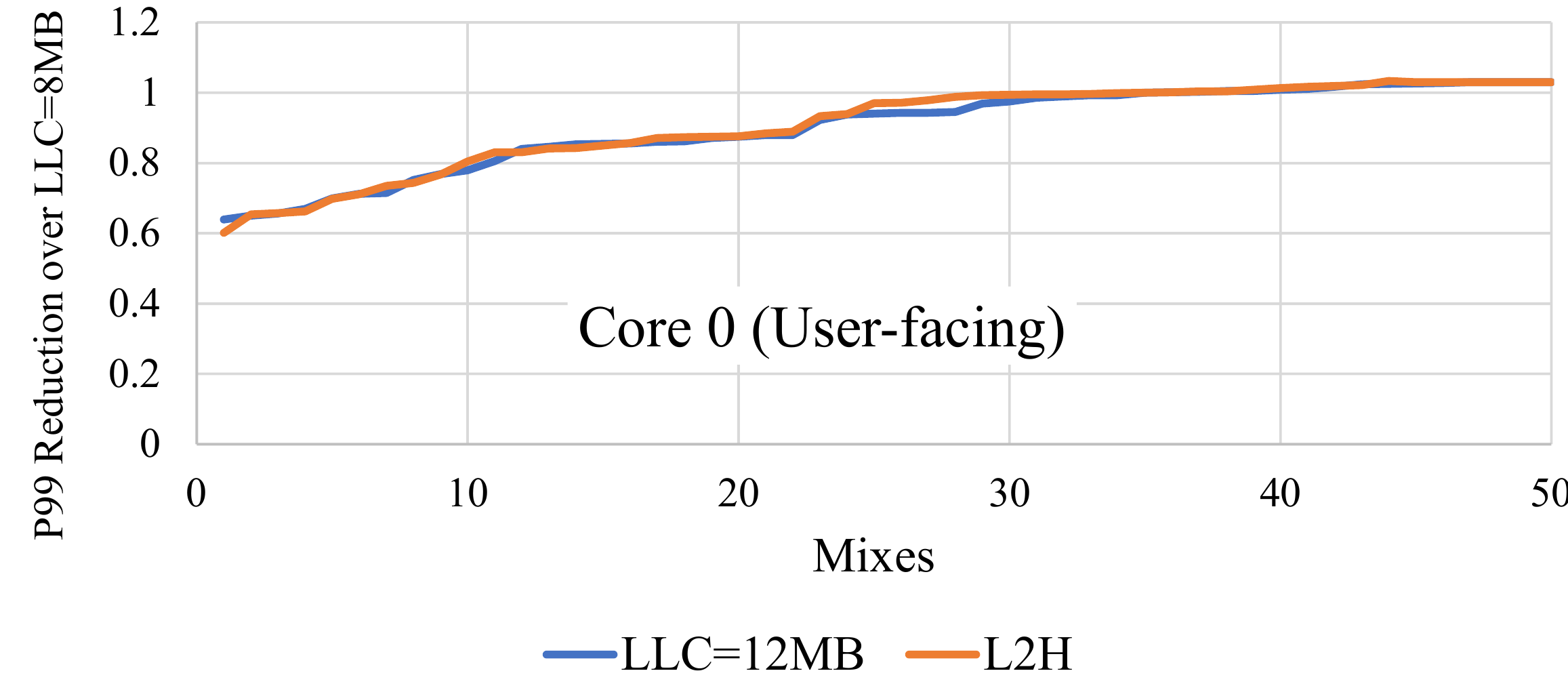}
  \includegraphics[width=3.2in]{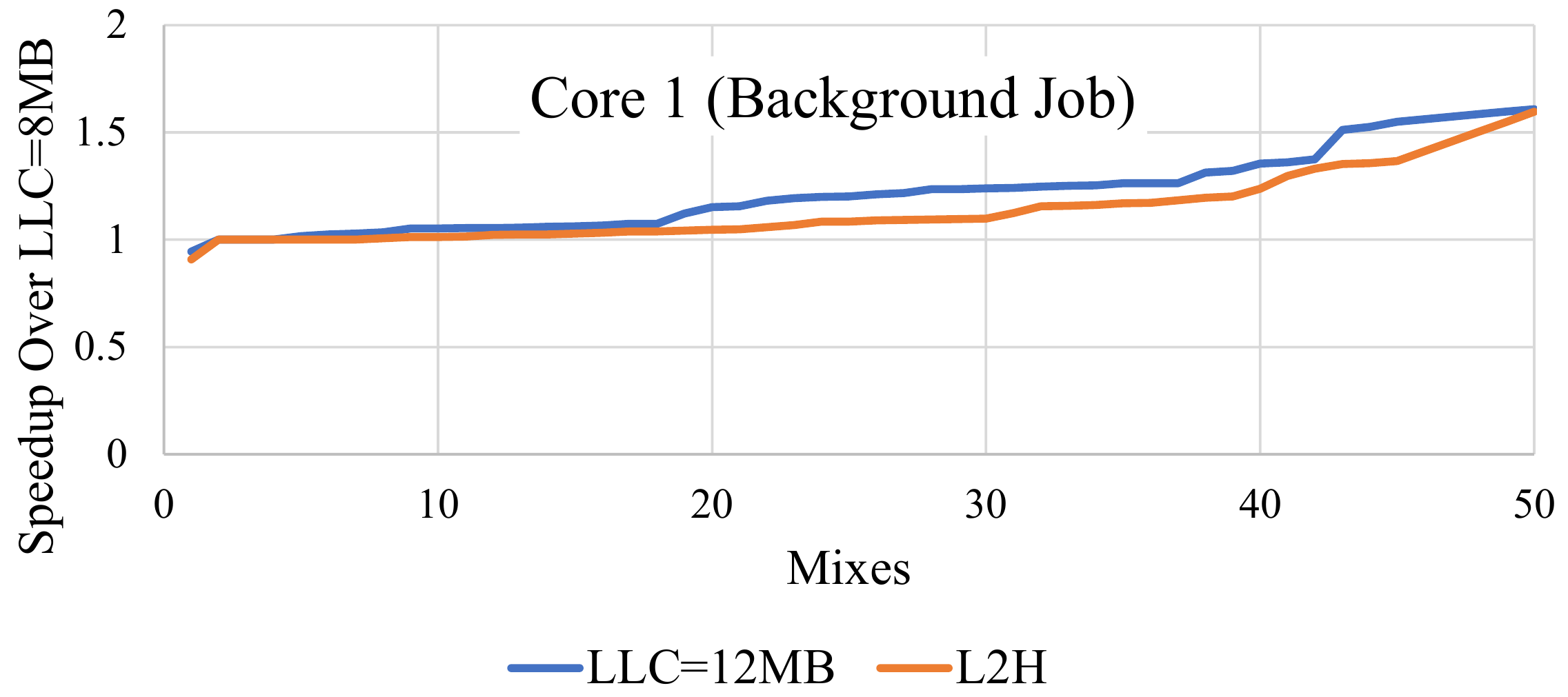}
  \caption{System performance s-curve, normalized to the baseline with an 8MB LLC. (Top) Normalized P99 latency of user-facing jobs; the lower, the better; (b) Background job speedup.}
  \label{fig:multi_s_curve}
\end{figure}

\noindent\textbf{Performance} We now focus on a more complex scenario, where there are two applications running: core 0 runs a user-facing application and core 1 runs a background job. Thus, there are two idle cores (50\% utilization). \fig{fig:multi_s_curve} shows the reduction in P99 for the user-facing application (top) and speedup for the background job (bottom). We normalize both to the baseline with an 8MB LLC. We sort the workloads in ascending order to yield S-curves. For the P99, the lower is the better, while for the speedup the higher is the better. We observe that P99 decreases to almost 60\%, while the background job is sped up by up to 50\%. We also show the 12MB LLC configuration. As can be seen, L2H closely follows the behavior of the larger 12MB LLC.

%-------------------- MULTICORE

\begin{figure}[t]
  \centering
  \includegraphics[width=3.0in]{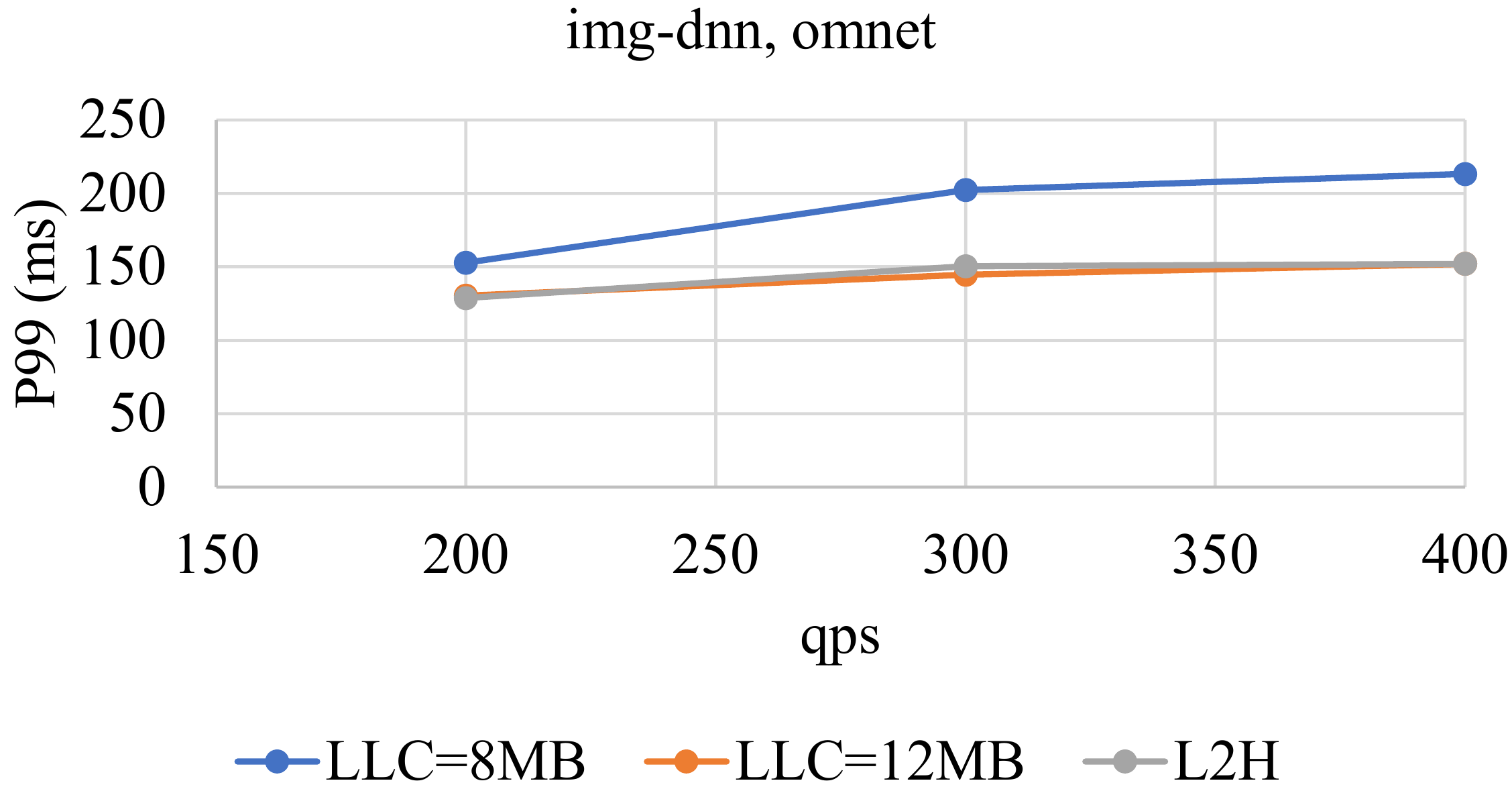}
  \includegraphics[width=3.0in]{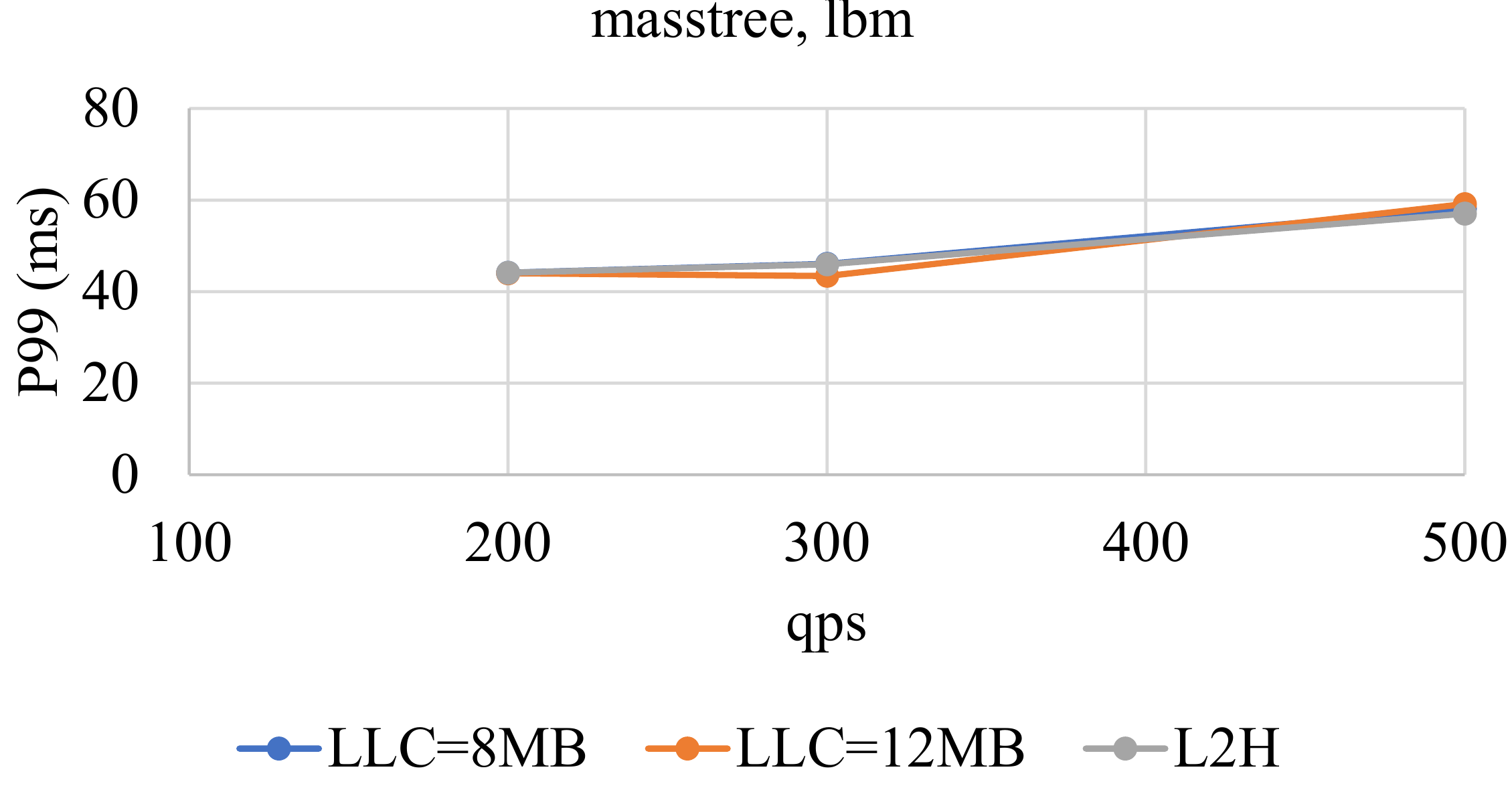}
  \caption{P99 latency of two pairs of appilications.}
  \label{fig:multi_examples}
\end{figure}

%-------------------- 

To better understand the results, we take a deeper look at two mixes shown in \fig{fig:multi_examples}. The first mix has \emph{img-dnn} as the user-facing job and \emph{omnet} as the background job. From the single-application experiments (\fig{fig:single_mpki}), we expect these two applications to be very sensitive to cache size. In this experiment, we vary the request rate from 200 to 400 qps. We make two observations: (1) as expected the absolute P99 latency increases compared to a single-application run (from 50ms to 126ms). However, the server is not saturated and (2) L2H helps P99 stay very close to that with the 12MB LLC. For example, at qps=400, the P99 latency for the baseline with an 8MB LLC is around 200ms while the L2H keeps it very close to that of the 12MB LLC at 150ms. This is significantly given that our result shows that \emph{omnet} IPC also improves by 4\% at the same time. It is evident that the load balancer has helped both applications to share the extra space provided by the idle cores.

\begin{figure}[t]
  \centering
  \includegraphics[width=\columnwidth]{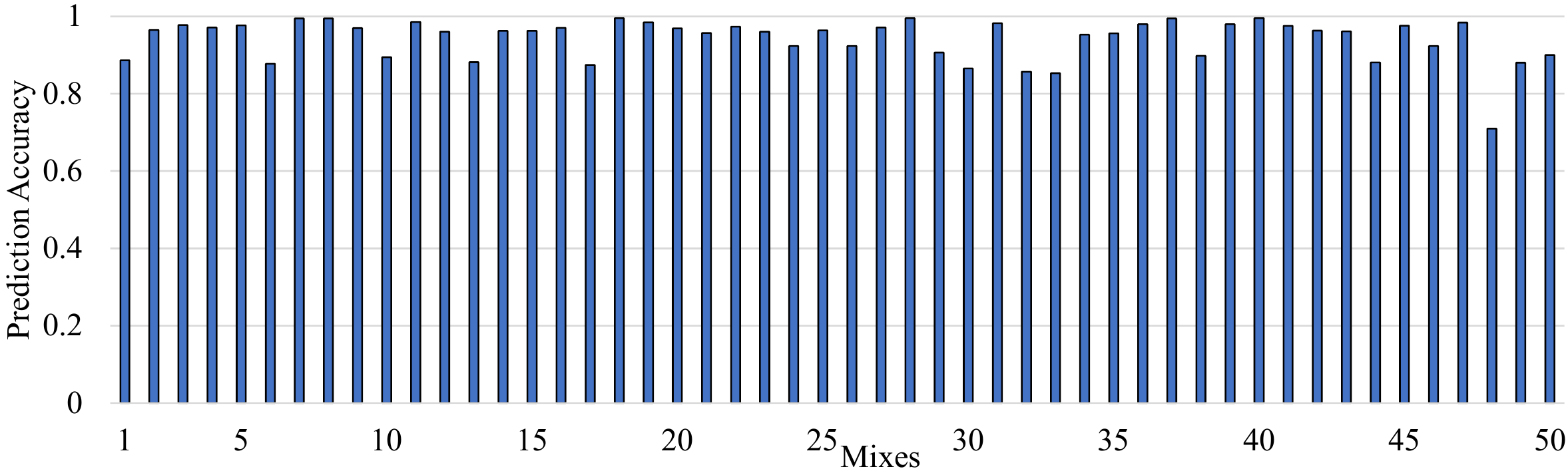}
  \caption{Multi-application prediction accuray.}
  \label{fig:multi_accuracy}
\end{figure}

\noindent\textbf{Prediction Accuracy} \fig{fig:multi_accuracy} shows the prediction accuracy for all 50 mixes. The average prediction accuracy is 86\% and ranges from 62\% to 99\%. Overall, the high prediction accuracy carries from the single-application experiments. We also measure how often the bloom filter is not warmed up and we need to refer or MPPP (15\%), the load is high and we must get the same output from both predictors (40\%), and finally how often we need one predictor to send a block up (45\%). We observe that all three situations are serviced well, given the high prediction accuracy.

\begin{figure}[t]
  \centering
  \includegraphics[width=\columnwidth]{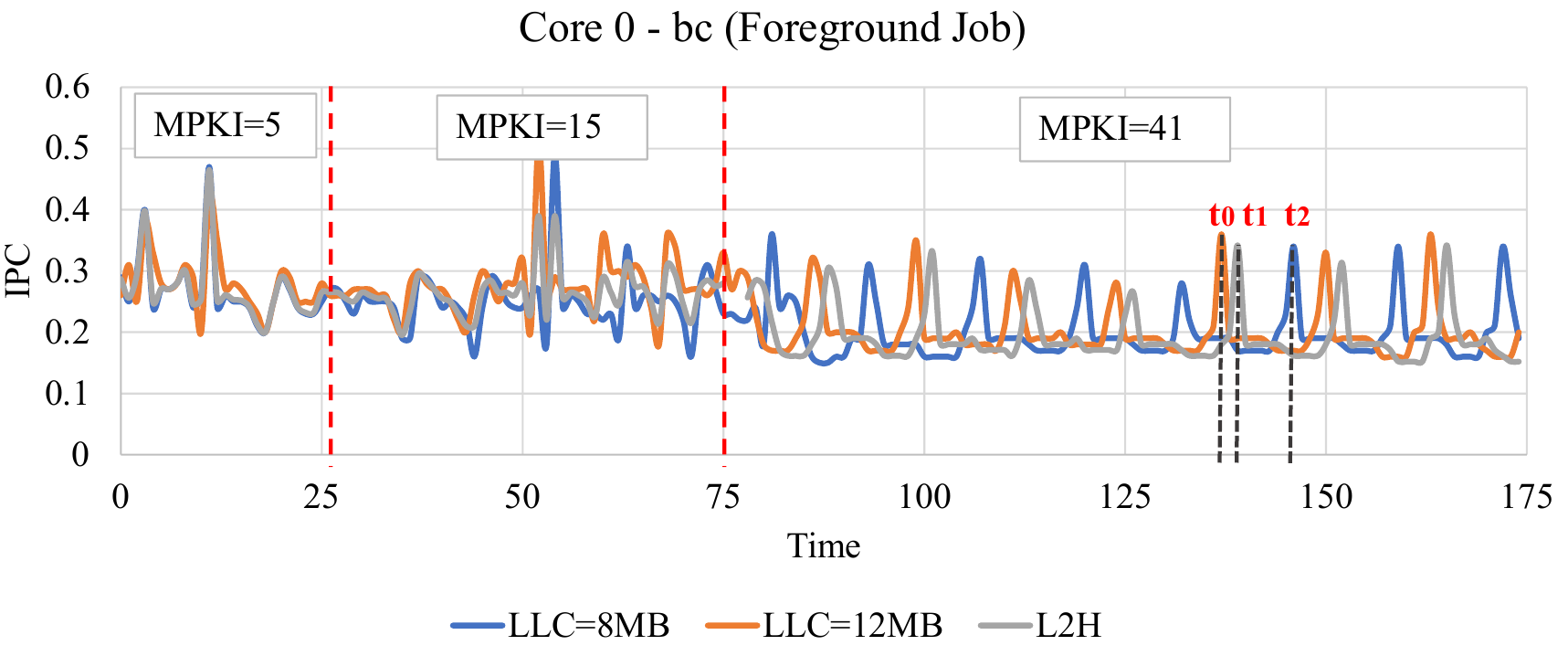}
  \includegraphics[width=\columnwidth]{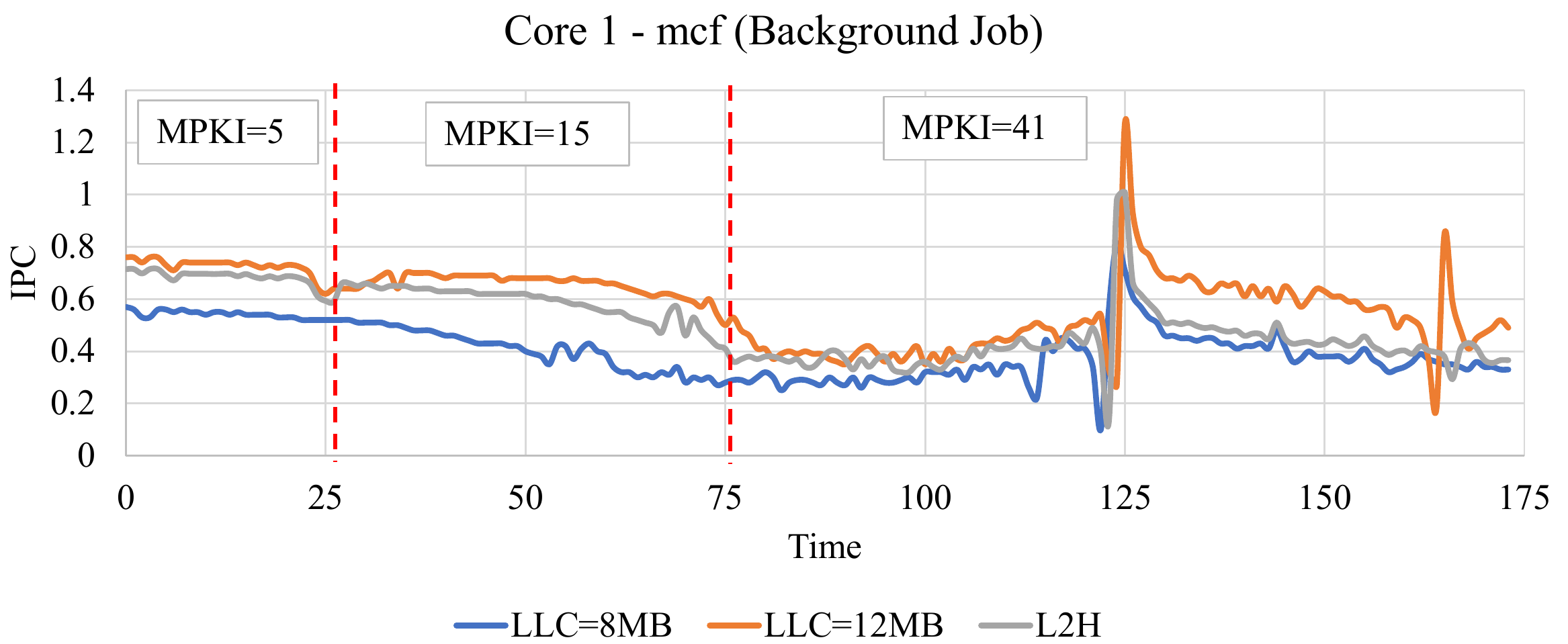}
  \caption{Load balancer analysis. When MPKI is low, L2H approaches the baseline with 12MB LLC as the foreground job yields extra space. When MPKI is high, the background job approaches the baseline with 8MB LLC as the load balancer gives extra space to the background job. }
  \label{fig:load_balancer_bc_mcf}
\end{figure}

\subsection{Load Balancer Analysis}
One major benefit of L2H is software transparency. The load balancer plays an important role to achieve this goal. To better understand how the load balancer works in practice, we designed a simple experiment where we varied the critical application MPKI to reveal how the load balancer works.

\fig{fig:load_balancer_bc_mcf} shows the absolute IPC for two applications and 3 systems (baseline with an 8MB LLC, a 12MB LLC, and L2H) over time: foreground job \emph{bc}, and background job \emph{mcf}. We pick \emph{bc} to be the foreground job because the input to this workload can be changed such that the MPKI changes. We call this workload foreground, and not user-facing because this is not a usual user-facing application. We could not find any Tailbnech applications whose MPKI changes easily. 
We use \emph{bc} with input $g19$ to have the foreground job show MPKI=5, input $u20$ to reach the MPKI to 15, and input $u21$ to increase the MPKI to 41. We annotate the figure to show these three MPKI regions over time. 

Based on the load balancer algorithm and \fig{fig:wchance}, we expect that in this first region (MPKI=5), the background job gets the majority of the extra space as the critical application has very low MPKI and is driven with a small graph ($0.95^5=0.77$ of alive \emph{mcf} blocks are sent up). We observe that in this region, all three system show very close IPC for \emph{bc}, and all provide enough cache for this application. Interestingly, for MPKI=5 and \emph{mcf}, we notice that L2H is very close to the baseline with 12MB LLC, and 13\% better than the baseline with 8MB LLC. Hence, the load balancer has redirected data blocks properly and fairly to private L2 caches in this region.

In the second region (MPKI=15), we expect that all \emph{bc} alive blocks and $0.95^{15}=0.46$ of \emph{mcf} alive blocks get the chance to stay on the chip because now the foreground MPKI has increases. We make two observations in this region: (1) \emph{bc} gets more space allowing it to follow the baseline with 21MB cache. Also, this extra space allows L2H and the 12 MB LLC to execute faster; the peaks are shifting to the right for the 8MB LLC; and (2) now the \emph{mcf} sits between the 8MB and 12MB LLCs because it now must yield the extra space. 

Finally, in the third region (MPKI=41), baseline the 12MB LLC and L2H continue to execute faster than baseline LLC=8MB for \emph{bc}. The difference between the peaks is now more visible; The peak at $t_0$ for the 12MB LLC arrives earlier than L2H ($t_1$), and the baseline 8MB ($t_2$). In this region, the background job approaches the baseline with 8MB LLC, mainly because the load balancer does not allow it to send the blocks up ($0.95^{41}=0.12$).

\subsection{Storage Overhead Analysis}
L2H uses two predictors. The bloom filter can store 4096 entries. It has 4 tables, each 4K, summing up to a total of 16KB storage overhead per processor. MPPP uses 256 sampled sets, adding up to 68.63KB. Other components in L2H are fairly small. Idle Core Map (ICM) and Critical Task Map each requires \emph{n} bits, where n is the number of cores (e.g., 128 bits = 16B). We store the sendup likelihood in a lookup table to avoid computation. This needs 100$\times$2B=200B storage. Overall, L2H needs 84.85KB storage for a 128-core processor.
%%% Local Variables:
%%% mode: latex
%%% TeX-master: "main"
%%% End:

\section{Related Work}
\label{sec:relatedwork}
The insight behind Morpheus \cite{morpheus} is similar to that of L2H, but for GPUs. The authors observe that increasing the number of SMs is not always useful and system performance stays constant after a certain number of SMs. They propose to not activate several SMs, and instead borrow some resources such as cache or register files from idle SMs. Apart from applying this idea to a different context (GPU vs. CPU in L2H), the differences are two-fold. First, idleness in L2H comes from natural underutilization in the cloud, while Morpheus needs to deactivate SMs to be able to borrow resources. This requires Morpheus to run profiling to find the optimal number of SMs for each application. Second, GPUs lack coherent caches, substantially increasing complexity and requiring extensive changes to the GPU microarchitecture. In contrast, L2H relies on existing mechanisms and adds off-the-critical path predictors at the LLC. Overall, both techniques address important underutilization scenarios, but very different ones.

Jenga \cite{jenga}, and Eva \cite{eva} address underutilization in caches by redesigning a new reconfigurable virtual cache hierarchy. Jenga defines a pool of caches where a run-time decides how each of them should be used. They propose an adaptive hierarchy allocation which finds the exact number of cache banks as well as the right cache level. They also propose a placement strategy called Bandwidth-aware data placement, where they try to put data blocks in the hierarchy where it makes more sense in terms of bandwidth. Jenga breaks the rigid hierarchy in the interest of reconfigurability, where L2H keeps the classic memory hierarchy but opens the path to use all levels automatically. Jenga requires OS and run-time support, while L2H is completely  transparent to software.

D2D \cite{d2d} split data hierarchy from metadata hierarchy allowing the data blocks to be found in the memory hierarchy with a single lookup. Separating metadata from data allows the authors to propose optimizations for data placement. However, D2D cannot utilize the unused cache, instead helps to find the block faster.

IBM Z16 \cite{ibmz16}, the latest generation of IBM mainframe processors, has 4 levels of caches L1=128KB, L2=32 MB, L3=up to 256 MB, and L4=2048 MB. L3 and L4 are called virtual caches similar to Jenga’s definition. They can be allocated on any of the share part of any L2 cache. Hence, with proper run-time management, the L2 waste can be reduced. For IBM z16 to work, the IBM Processor Resource/Systems Manager (PR/SM) scheduler and the z/OS WLM and dispatcher must work together to enable and use the large caches. IBM also optimizes the lithography to reduce the cache access latency. Z16 also needs a translation layer to be able to find the data block in banked caches scattered throughout the chip. We believe that the classic hierarchy offers a simpler design, and can be fixed to make better use of the caches with L2H.

CATCH \cite{catch} proposes a criticality-aware tiered cache hierarchy, where the authors argue that having a large L2 cache is not an efficient design choice as L2 is not large enough to capture the working set completely, nor as fast as the L1 cache. Instead, CATCH proposes to remove the L2 cache and compensate for its loss with new inter-level prefetchers to move data in a timely manner between a larger LLC and the L1 caches. We argue that the L2 is still very valuable. First, it is very effective for some applications~\cite{lp}. Second, L2 is very effective in reducing the number of coherence requests as it is usually inclusive of L1 cache. Thus, keeping L2 is a good design choice, and its low hit ratio can be compensated for by borrowing/lending space from/to neighboring cores.

Dead block prediction is another way to increase LLC utilization. A cache block is dead if it has exhausted its useful lifetime in the cache, and can be evicted to make space for other blocks \cite{mp, PercDeadblock,sampling}. Using perceptron-based prediction proposed in some prior work  \cite{mp,PercDeadblock }. Using sampling to detect dead blocks suggested by authors of \cite{sampling}.

%Designing customized caches to accelerate the address translation proposed in \cite{single_walk}, or GNNCache proposes a new cache design customized for graph neural networks \cite{GNNCache}. Replacement policy can increase cache utilization by early evicting blocks or promoting alive blocks \cite{RP1, hawkeye, MLRL,min_belady}. Dynamic Non-uniform cache access (DNUCA) architectures is a class of solutions where a data block is placed onto an LLC bank that is predicted to be closer to consumer core \cite{nuca1,nuca2,nuca3,nuca4,nuca5,nuca6,nuca7,nuca8,nuca9,nuca10,nuca11}.

%CAT
Cache partitioning is a strategy to provide quality of service for over-provisioning datacenters cores \cite{swap,dirigent,heracles,dcaps,satori,pagecolor,rhythm,selfa,pons,clite,copart,provisioning}. They use Intel Cache Allocation Technology to partition LLC on a real machine or a cluster of machines. 
%They use heuristics such as page coloring, and slowdown estimation to provide fairness and quality of service at different levels (applications, tasks, kernel, etc) \cite{kpart,swap,dirigent,tako,pagecolor,rhythm,selfa,pons,clite,copart,provisioning, swap}. Using machine learning to partition ways is also proposed in SATORI \cite{satori}. 
L2H is orthogonal to cache partitioning, although try to provide fairness for datacenter applications.

\section{Conclusion}
\label{sec:conclusion}
We propose L2 Harvester (L2H), a simple approach to harvest unused L2 caches in low-utilization beefy server processors. We make this observation that number of cores and cache sizes (both L2 and LLC) are constantly increasing while the core utilization struggles to catch up in public clouds (mostly $<$40\% in Azure, and around 20-50\% in Alibaba). To address this shortcoming, we devise a mechanism to detect LLC evictions that are not dead, and redirect them to upper L2 caches, if the system load permits. L2H is implemented with minimal changes to the current architecture. Our experimental results show that L2H improves system performance by up to 2$\times$, and 32\% for single-application and multiple-application, respectively.

\section{ACKNOWLEDGMENTS}
We acknowledge Chameleon@TACC for providing computation resources. This work was supported by National Science Foundation under Grant
$\#1719061$

%\clearpage
%%%%%%%%% -- BIB STYLE AND FILE -- %%%%%%%%
\bibliographystyle{IEEEtranS}
\bibliography{./references}
%%%%%%%%%%%%%%%%%%%%%%%%%%%%%%%%%%%%
\end{document}